\documentclass[a4paper,12pt]{article}
\pdfoutput=1
\usepackage{amsmath,amsfonts,amssymb,epsfig}
\usepackage{slashed}
\usepackage{cite}
\usepackage[width=0.9\textwidth,margin={30pt,30pt},font=small,labelfont=bf]{caption}
\numberwithin{equation}{section}
\usepackage[utf8]{inputenc}
\usepackage{hyperref}
\usepackage{subfig}

\usepackage{xcolor}

\hypersetup{
	bookmarks=true,         
	unicode=true,          
	pdftoolbar=true,        
	pdfmenubar=true,        
	pdffitwindow=false,     
	pdfstartview={FitH},    
	pdftitle={GeneralisedMEP },    
	pdfauthor={Author},     
	pdfsubject={Subject},   
	pdfcreator={LD, JJ, ML},   
	pdfproducer={Producer}, 
	pdfkeywords={MEP} {Tunneling} {QFT} {Bubble} {Membrane} {Phase transition}, 
	pdfnewwindow=true,      
	colorlinks,
linkcolor={red!0!black},
citecolor={blue!50!black}, 
urlcolor={blue!80!black}
}
\usepackage{graphicx}
\usepackage{cite}
\usepackage{color}
\usepackage{bm}
\usepackage{multirow}

\usepackage{xspace}

\newcommand{\scr}[1]{\ensuremath{\mathcal{#1}}}

\def\nn{\nonumber}

\def\d{\partial}

\def\vep{\varepsilon}

\def\out{\mathrm{out}}
\def\ins{\mathrm{in}}

\def\phiout{ \varphi_{ \text{\tiny out}}}
\def\phiin{ \varphi_{ \text{\tiny in}}}

\def\psif{\Psi [\phi]}
\def\mlam{m_\lambda}

\def\intv{\int_{\scr{V}}}

\newcommand{\fd}[2]{\ensuremath{\frac{\delta #1}{\delta #2}}}
\newcommand{\fdd}[2]{\ensuremath{\frac{\delta^2 #1}{\delta #2{}^2}}}
\def\fs{\phi_s}
\def\dphip{h_\perp}

\def\dell{\d_\ell}
\def\dlam{\d_\lambda}

\def\nn{\nonumber}

\normalsize
\newcommand\blfootnote[1]{%
	\begingroup
	\renewcommand\thefootnote{}\footnote{#1}%
	\addtocounter{footnote}{-1}%
	\endgroup
}

\begin{document}
	
	\title{\LARGE \vspace{-1cm} {\bf Generalised escape paths \\[0.3em] for dynamical tunneling in QFT}}
	
	\author{\\[2em]  Luc Darm\'e,$^{1,a}$\blfootnote{$^a$  \url{luc.darme@ncbj.gov.pl}} Joerg Jaeckel$^{2,b}$\blfootnote{ $^b$ \url{jjaeckel@thphys.uni-heidelberg.de}}  \let\theblfootnote\relax and  Marek Lewicki$^{3,4,c}$\blfootnote{$^c$ \url{marek.lewicki@kcl.ac.uk}}\\[5ex]
		\small {\em $^1$ National Centre for Nuclear Research, }\\ 
		\small {\em ul. Pasteura 7, 02-093 Warsaw, Poland }\\
		\small {\em $^2$ Institut f\"ur Theoretische Physik, Universit\"at Heidelberg,}\\
		\small {\em Philosophenweg 16, 69120 Heidelberg, Germany,}\\
		\small {\em $^3$ King's College London,}\\
		\small { \em Strand, London, WC2R 2LS, United Kingdom,  }\\
		\small {\em $^4$ Faculty of Physics, University of Warsaw}\\
		\small{\em  ul.\ Pasteura 5, 02-093 Warsaw, Poland}\\
	}
	%
	\date{}
	\maketitle

	\abstract{We present a formalism based on the functional Schr\"odinger equation to analyse time-dependent tunneling in quantum field theory at the semi-classical level. The full problem is reduced step by step to a finite dimensional quantum mechanical setup and solved using the WKB approximation. As an example, we consider tunneling from a homogeneous oscillating initial state in scalar quantum field theory.
		
		\vspace{1cm}
		
	}
	\newpage 
	
	\tableofcontents
	
	\setcounter{footnote}{0}
	
\section{Introduction}

The tunneling of a scalar field settled in a local minimum of its potential to its global minimum is a fascinating process which appeals more to the quantum aspects of Quantum Field Theory (QFT) than to its classical ones. Furthermore, far from being a purely theoretical problem, tunneling in QFT is a critical ingredient in many current areas of research including, for instance, the study of phase transition in the early universe and gravitational wave production, vacuum stability of the electroweak vacuum or baryogenesis.

In a seminal paper of Coleman~\cite{Coleman:1977py} it was understood that the tunneling rate could be obtained by studying the so-called ``Most Probable Escape Path'' (MPEP) introduced in~\cite{banks_coupled_1973,banks_coupled_1973-1}. This path is the trajectory in the field space along which the tunneling probability is maximal and it directly corresponds to a solution of the equations of motion in Euclidean time. This result together with subsequent estimation of first-order quantum corrections to this path by Coleman and Callan in~\cite{Callan:1977pt,Coleman:1980aw} led to the well-known formula for the tunneling rate per unit of volume $\Gamma / V$ in the thin-wall approximation,
\begin{equation}
\frac{\Gamma}{V} = A e^{\displaystyle 27 \pi^2 \frac{\sigma_0^4}{2 \epsilon^3}} \ ,
\end{equation}
where $\sigma_0$ and $\epsilon$ are respectively the tension and vacuum energy of the bubble whose precise expression will be given in Sec.~\ref{sec:Oscillating}, and $A$ is a quantum correction coefficient which was first estimated in~\cite{Callan:1977pt}. In parallel it was realised in~\cite{Gervais:1977nv,Bitar:1978vx} that a similar result could be obtained using the Functional Schr\"odinger Equation (FSE) along the same MPEP.  While being a functional equation, the FSE is usually solved in the semi-classical limit by reducing the field evolution to a given path, out of which the wave functional is suppressed by factor of $\hbar$. This reduces the problem to a one-dimensional quantum mechanical one which is readily solved. 

However, both approaches fail to describe the case of time-dependent tunneling, either because the rotation to imaginary time performed by~\cite{Coleman:1977py} is no longer appropriate, or, in the FSE case because the reduction to a single path no longer properly describes the dynamics of the system. Interestingly, the last decade has seen a resurgence of interest in generalisation of the imaginary time rotation of~\cite{Coleman:1977py} to a more generic complex time trajectory, with important successes in describing tunneling with some initial dynamics in Quantum Mechanics (QM)~\cite{Turok:2013dfa,Cherman:2014sba,Bramberger:2016yog,Anastopoulos:2017fgw}. Let us also note that recent results using a real time formalism (potentially coupled to a lattice simulation) are very promising~\cite{Braden:2018tky,Hertzberg:2019wgx,Ai:2019fri} and could in principle be extended to our dynamical initial state problem.
Similarly it might be interesting to see whether the significant simplifications in the calculation of tunneling actions based on generalized potentials~\cite{Espinosa:2018hue,Espinosa:2018szu} could be applied to time dependent tunneling problems.

While application of these techniques to the QFT case in presence of a dynamical initial field is often referred to as a possible extension, to the best of our knowledge no definite progress has been made. On a parallel but related topics, there have been new developments in the case where the dynamics is provided not by the field itself, but rather by the potential which possesses a non-negligible time-dependence using both the instanton and the FSE approach~\cite{Widrow:1991xu,Simon:2009nb} (as may be relevant, e.g. in cosmological setups). A first attempt at describing the case of a dynamical initial state for vacuum tunneling in the membrane approach has been made by~\cite{KeskiVakkuri:1996gn}. While we tackle the same problem in Sec.~\ref{sec:Oscillating}, our conclusions differ from~\cite{KeskiVakkuri:1996gn}. A more detailed comparison of both approaches is made in Sec.~\ref{sec:Oscillating}, but the essential point is that the oscillations of the field should not be treated as a background ``potential''  to the tunneling process, but rather included as initial state of the tunneling, with the tunneling path starting from this oscillating state.

In this work, will build on the FSE formalism introduced in~\cite{Gervais:1977nv,Bitar:1978vx,Widrow:1991xu} to describe tunneling from a dynamical state in QFT in the semi-classical limit. Our key suggestion is that such process should be described by reducing the full QFT problem to \textit{multi-dimensional} quantum mechanical one, instead of the one-dimensional approach used in static tunneling problems. Once this reduction has been performed, the system resembles the problem of multi-dimensional tunneling in Quantum Mechanics, as described in~\cite{Bowcock:1991dr}, allowing to solve for a time-dependent tunneling rate.

We start Sec.~\ref{sec:2Dred} with the reduction of the FSE on a two-dimensional sub-space. This is complemented by a discussion of how the time-dependent problem can be approached using the time-independent version of the FSE. Sec.~\ref{sec:Oscillating} illustrates the procedure developed in Sec.~\ref{sec:2Dred} for the case of the vacuum decay of an oscillating initial state thereby improving on the heuristic results obtained in~\cite{Darme:2017wvu} both analytically and numerically. The results are briefly discussed in Sec.~\ref{sec:conclusions}.
Appendices~\ref{sec:tunQM} and~\ref{app:1dRed} review basics of the WKB approximation as well as the usual one-dimensional reduction of the FSE for ordinary tunneling.
Appendix~\ref{sec:Twotunneling} briefly ventures into the possibility to describe two successive decoupled tunneling events using the same formalism.

\section{Semi-classical approach to dynamical tunneling in QFT}
\label{sec:2Dred}

Our formalism is based on the Functional Schr\"odinger Equation (FSE), which describes the time evolution of the wave functional $\Psi [\phi,t]$ given a potential $V(\phi)$ (see, e.g., chapter 10 of \cite{Hatfield:1992rz}).
\begin{align}
\label{eq:timeFSE}
i \hbar   \frac{\d}{\d t} \Psi [\phi,t] = -\frac{\hbar^2}{2} \intv dx^3 \frac{\delta ^2}{\delta \phi ^2} \Psi [\phi,t] + U(\phi)  \Psi [\phi,t] \ ,
\end{align}
where $\scr{V}$ is a control volume and we have introduced the effective potential
\begin{align}
\label{eq:effpot}
U(\phi) = \intv dx^3  \left[\frac{(\nabla \phi )^2}{2} + V(\phi) \right] \ ,
\end{align}
which includes the spatial gradient of the field $\phi$. 
As for the quantum mechanical Schr\"odinger equation, the norm of the wave functional $\Psi[\phi]$ gives a measure of the likelihood of the occurrence of the field configuration $\phi$. 

\bigskip

In this section we discuss the essential steps to obtain our approximate solution to the time-dependent problem. We start with the FSE which we approach with a suitable WKB ansatz.
In a first step we then simplify the time-dependent functional equation to an ordinary quantum mechanical Schr\"odinger equation by considering a suitable two-dimensional sub-space of all the fields. The two dimensional sub-space is then determined by an appropriate ``equation of motion'' for the fields. This is effectively the two-dimensional generalization of the one dimensional Maximum Probability Escape Path used to describe tunneling from a time-independent meta-stable state. The two-dimensional generalization allows us to match the tunneling solution also at times when the time-derivative of the field is non-vanishing.

Even in this simplified form solving the full time-dependent problem seems impractical. We therefore reduce the problem to a time-independent one by considering energy eigenstates.
This allows us to further simplify and obtain concise equations of motion for the semi-classical field solutions. These can then be solved with suitable boundary conditions for the initial field value and its time derivative. We do this for a concrete example in Sec.~\ref{sec:Oscillating}.

However, considering energy eigenstates is a non-trivial step, as such states are quasi-static. To motivate our procedure we recall the construction of  coherent states which are close to our oscillating classical field solutions. We then argue that in the saddle point approximation matching conditions should indeed focus on the ``classical'' field values, i.e. the expectation  values.
We introduce wave-packet solutions in order to focus on a time-independent quantum mechanical problem and show how to relate the tunneling rate for the time-dependent case to this time-independent formalism. 

We restore explicit factors of $\hbar$ whenever useful.

\subsection{From the  functional Schr\"odinger equation to quantum mechanics}\label{sec:FSE}

A first important comment is that, similarly to its QM counterpart, the FSE describes the behaviour of a quantum field system in isolation. All space integrals are therefore taken on a certain control volume $\scr{V}$ and we will have to assume that the inside of the control volume is isolated on the time-scale relevant  to the processes considered. For the particular case of bubble nucleation considered below, we will assume the control volume to be a few times bigger than the radius of the vacuum-to-vacuum bubble.\footnote{The precise definition of the control volume as well as the validation of this assumption will rely on estimating the effect of quantum decoherence. While in the following we will neglect decoherence and assume that the system is perfectly isolated, it seems clear that this issue should be studied carefully in further work.}

Following the intuition of~\cite{banks_coupled_1973,banks_coupled_1973-1}, we will use a ``saddle point approximation'' in that we will try to evaluate this equation only along a particular field hypersurface $\scr{H}$ (typically of dimension one or two) $\phi_{s}(\vec{x},\lambda^i)$ parametrised by $\lambda^i$, such that $\phi_s$ is a saddle point of the wave functional $\psif$,
\begin{align}
\label{eq:orthH}
\left.\frac{\delta }{\delta \phi_\perp } \psif \right|_{\phi_s} = 0 \ ,
\end{align}
where we have labelled by $\phi_\perp$ all the fields configuration orthogonal to the hypersurface $\scr{H}$.

For later use we note that any perturbation $\dphip$ orthogonal to the surface $\scr{H}$  satisfies (see e.g., Sec 5. of~\cite{Bitar:1978vx}),
\begin{align}
\label{eq:perpphi}
\d_\lambda \fs \cdot \dphip = \int dx^3 \d_\lambda \fs (x) \dphip (x) = 0 .
\end{align}

\bigskip
The main difference to the standard approach (reviewed briefly in Appendix~\ref{app:1dRed}) is that we will consider a two-dimensional subspace instead of the usual one-dimensional MPEP. The reason is that the one-dimensional setup allows matching of the classical to the quantum regions only at the turning point of the classical motion, where the momentum of an incoming wave vanishes. For a time-dependent classical solution, this only occurs at some particular times.
To allow more general matching conditions a more flexible, general approach is needed.
In this section we discuss how to use this idea in practice and therefore to reduce the FSE on a multi-dimensional (in particular we consider the two dimensional case) hypersurface. 
Along such a hypersurface, the problem can be reduced to a tractable case of multi-dimensional quantum mechanics as studied in~\cite{Bowcock:1991dr}. In the following we concentrate only on the leading order in $\hbar$ dependence of the wave functional on this hypersurface.

\bigskip

In the following, we will suppose that the wave functional 
takes the form
\begin{align}
\Psi = e^{\frac{i}{\hbar} ( F + i G )}  \ .
\end{align}
The difficulty of the procedure is that the precise shape of the hypersurface $\scr{H}$ depends on $\Psi$ and will be ultimately fixed by solving~\eqref{eq:perpphi}. 

Introducing this ansatz for the solution in~\eqref{eq:timeFSE} and decomposing between real and imaginary part, the FSE becomes  
\begin{align}
\label{eq:FSE2}
& \intv dx^3 \left[ -\left(\fd{F}{\phi}\right)^2 + \left(\fd{G}{\phi}\right)^2 - \hbar \fdd{G}{\phi} \right] = 2 \d_t F + 2 U(\phi) \\
& \intv dx^3 \left[ 2\fd{F}{\phi}  \fd{G}{\phi} - \hbar \fdd{F}{\phi} \right] =  -2 \d_t G  \ \nn .
\end{align}

We aim at reducing the FSE~\eqref{eq:FSE2} on a surface $\scr{H}$ in field space given by the field configurations $\phi_s(\lambda^i)$, such that 
\begin{align}
\label{eq:firstEoM}
\displaystyle \frac{\delta }{\delta \phi_\perp }\Psi[\phi] \bigg|_\scr{H} =  0  \qquad \Rightarrow  \qquad \begin{cases}
\displaystyle \frac{\delta F}{\delta \phi_\perp } \bigg|_\scr{H} =  0  \\
\displaystyle \frac{\delta G}{\delta \phi_\perp } \bigg|_\scr{H} =  0 
\end{cases}\ .
\end{align}
Note that at this point the $\lambda^i$ are simply parameters along the hyperplane $\scr{H}$. Using~\eqref{eq:firstEoM} (as, e.g., in~\cite{Bitar:1978vx,copeland_no_2008}) we have in particular that on this hyperplane,
\begin{align}
\d_i G =  \intv  \fd{G}{\phi} \d_i \phi_s  \nn \\
\d_i F =  \intv  \fd{F}{\phi} \d_i \phi_s  \nn \ .
\end{align}
Specifying to the two-dimensional case and writing $\lambda^i = (\ell,\lambda)$, we can define a metric $g$ on the surface $\scr{H}$  given by 
\begin{align}
g_{ij} &\equiv \displaystyle \begin{pmatrix}
m_\ell &  X_{\ell\lambda}  \\[0.4em] 
X_{\ell\lambda}  & \mlam
\end{pmatrix} \\[0.6em]
g^{ij} &= \displaystyle \frac{m_\ell \mlam}{m_\ell \mlam-X_{\ell\lambda}^2} \begin{pmatrix} \displaystyle
\frac{1}{m_\ell} &  -\displaystyle\frac{X_{\ell\lambda}}{ m_\ell \mlam} \\[0.8em]  
-\displaystyle \frac{X_{\ell\lambda}}{ m_\ell \mlam}  & \displaystyle\frac{1}{\mlam}
\end{pmatrix} \ \nn .
\end{align}
where we have introduced the normalisation for the field 
\begin{align}
\label{eq:masses}
m_\ell &\equiv \intv dx^3 (\d_\ell \phi_s)^2  \qquad \qquad \mlam \equiv \intv dx^3 (\d_\lambda \phi_s)^2  \  ,
\end{align}
and the cross-product
\begin{align}
X_{\ell\lambda} \equiv  \intv \d_\ell \phi_s \d_\lambda \phi_s \ .
\end{align}
Using this notation and the usual Einstein indices summation convention, the functional derivative squared term in Eq.~\eqref{eq:FSE2} can now be written as
\begin{align}
\intv dx^3 \left[ -\left(\fd{F}{\phi}\right)^2 + \left(\fd{G}{\phi}\right)^2 \right] = \d^i G \d_i G - \d^i F \d_i F \ .
\end{align} 

Finally, we use the WKB expansion in which the semi-classical limit is taken by considering the decomposition 
\begin{align*}
F &=  \sum_{n=0}^\infty \hbar^n F_n \\
G&=  \sum_{n=0}^\infty \hbar^n G_n \ .
\end{align*} 
In the following we are going to concentrate on the zeroth order terms since our primary objective is to obtain the exponent of the tunneling rate. A short review of the WKB approximation in quantum mechanic is provided in Appendix~\ref{sec:tunQM}. In particular, we need to ensure the hierarchy,
\begin{eqnarray}
\left(\fd{F_0}{\phi}\right)^2 - \left(\fd{G_0}{\phi}\right)^2 & =& \d^i G_0 \d_i G_0 - \d^i F_0 \d_i F_0 ~\gg~ \hbar \fdd{G_0}{\phi} \\
\fd{F_0}{\phi} \fd{G_0}{\phi} &\gg&  \hbar \fdd{F_0}{\phi} \ ,
\end{eqnarray}
where the first equality can be  derived from~\eqref{eq:perpphi}. We thus recover the usual fact that the WKB approximation breaks down at the boundaries where the ``momentum'' vanishes.\footnote{In quantum mechanics, the system behaviour around the boundaries can be easily described using Airy functions. This leads to connections formulas describing the phase shifts across the interfaces and determining the pre-exponential factor. In the following we will focus only on the exponential part, for which the detailed treatment is not necessary.} We obtain the reduced set of equations
\begin{align}
\label{eq:FSE_time2}
&\d^i G \d_i G - \d^i F \d_i F = 2 (U+\d_t F)  \\[0.7em]
&\d^i F \d_i G = -\d_t G \ .
\end{align}

We thus obtain a formulation similar to the one used in~\cite{Bowcock:1991dr} to describe QM tunneling in a multi-dimensional potential. However, the mass parameters are now replaced by the inverse metric $g^{ij}$. We will focus in the rest of the paper on the case of stationary problems, for which the scalar potential $V[\phi]$ is time-independent and the time derivatives can be replaced by the energy of the initial state.

\subsection{Time-independent system and equation of motion}
\label{Sec:multiRed}

We can now focus on the time-independent system of equations assuming an energy eigenstate, $\Psi[\phi,t]\sim \exp(-iEt/\hbar)\Psi[\phi]$. Then we have,
\begin{equation}
\partial_{t}F=-E \, .
\end{equation}
Using this the FSE~\eqref{eq:FSE2} takes the simpler form,
\begin{align}
\label{eq:FSE3}
&\d^i G \d_i G - \d^i F \d_i F = 2 (U-E)  \nn \\[0.6em]
&\d^i F \d_i G = 0 \, .
\end{align}
This needs to be completed by the yet-unsolved equation of motion for the field configuration $\phi_s$, Eq.~\eqref{eq:firstEoM}. 

These are exactly the equations we would expect from the Schr\"odinger equation governing multi-dimensional tunneling. This equation can in principle be solved perturbatively using a generalised version of the procedure of~\cite{Bowcock:1991dr}. However, there are two additional complications. First the two-dimensional space is curved with metric $g_{ij}$. Second the field configuration $\phi_s$ that must be a solution of the functional equation~\eqref{eq:firstEoM} is not known a priori and consequently, neither is $U$.

A useful ansatz is to write Eq.~\eqref{eq:FSE3} in the form,
\begin{align}
\label{eq:Glines}
g^{ij} \d_i G  \d_j G = f \, , 
\end{align}
where $f$ is a function of the coordinates on $\scr{H}$ that does not depend on $G$ itself,
\begin{equation}
\label{eq:fdef}
f = 2 (U-E) + g^{ij} \d_i F \d_j F.
\end{equation}
This then can be solved using a ``momentum transfer method''. To implement this, we introduce the vector $k^i$ defined by,
\begin{align}
k^i = g^{ij} \d_j G \, ,
\end{align}
and look for the integral curve (G-lines) of $k^i$. Along those, Eq.~\eqref{eq:Glines} reduces to
\begin{align}
k^i \d_i G  = \d_s G  = f \, ,
\end{align}
where we have parametrised the position on the G-line by $s$, such that $ k^i \d_i \equiv \d_s$ on the G-line. The G-lines are then simply found by introducing a coordinate vector $X^i (s) = (\ell(s),\lambda(s))$ such that $\d_s X^i = k^i$. Noting $\nabla$ and $\Gamma^a_{\phantom{a} bc}$ the usual covariant derivative and Christoffel symbols for the metric $g_{ij}$, we then derive from Eq.~\eqref{eq:Glines} \footnote{Notice that a more symmetric form in $G$ and $F$ is 
	\begin{align}
	k^i \nabla_i k^j -  l^i \nabla_i l^j = \d^j U
	\end{align}
	where we have introduced the vector $l^i$ along the F-line as $l^i = g^{ij} \d_j F$. We further have trivially $k^i l_i = 0$ and $k^i k_i - l^i l_i = 2(U-E)$ from~\eqref{eq:FSE3}.
}
\begin{align}
2 g^{ij} \nabla_k (\d_i G) k_j = \nabla_k f \, .
\end{align}
After a bit of algebra, we obtain the parametric equation for the G-lines,
\begin{align}
\label{eq:FullGline}
\d_s^2 X^a + \Gamma^a_{\phantom{a} bc} \d_s X^b \d_s X^c = \frac{1}{2} g^{ab} \d_b f \, .
\end{align}
This should be solved using~\eqref{eq:Glines} written as,
\begin{align}
\label{eq:Glines2}
g_{ab}  ~\d_s X^a \d_s X^b= f \, , 
\end{align}
to obtain the initial values for $\d_s X^b$. This last equation should be seen as a type of ``energy conservation'' along the G-lines, notice in particular how $m_{\ell}$ and $m_{\lambda}$ from Eq.~\eqref{eq:masses} do indeed play the role of masses since the left-hand side of~\eqref{eq:Glines2} looks like a kinetic energy term.

Along any such line $G(s_f)$ takes the simple form
\begin{align}
G = \int_{s_i}^{s_f} ds f (s)   = \int_{\lambda_i}^{\lambda_f} d \lambda \sqrt{ \mlam f} \, .   
\end{align}
Here, $s_i$, $s_f$, $\lambda_i$ and $\lambda_f$ denote initial and final values of the respective parameters.
In the second equality we have restored the arbitrariness of the parametrisation of the G-line by replacing the parameter $s$ (with norm $\sqrt{m_s} = \sqrt{f} $) by a generic parameter $\lambda$ (with norm  $\sqrt{\mlam}$ ). We can now solve the saddle point equation~\eqref{eq:firstEoM} by considering a small orthogonal perturbation $\dphip$ around $\fs$ and ensuring that $\delta G = 0$ with respect to field configuration parametrised by $\lambda$, as in, e.g.~\cite{Bitar:1978vx}. We obtain, on the hyperplane $\scr{H}$,
\begin{align}
\label{eq:perpphi2}
\delta{G}_\perp = G (\fs + \dphip) - G (\fs )= 0 &~\Rightarrow~ \int_{\lambda_i}^{\lambda_f} d \lambda ~\delta \left(\sqrt{ \mlam f} \right)  = 0 \nn \\
&~\Rightarrow~ \int_{\lambda_i}^{\lambda_f} d \lambda  \left( \sqrt{\frac{f}{\mlam}} \delta{ \mlam }+\sqrt{\frac{\mlam}{f}} \delta{f} \right) = 0 \, .
\end{align}
It is clear that one of the simplest choice for the parameter $\lambda$ is such that 
\begin{align}
\label{eq:gaugefix}
m_{\lambda} = f
\end{align}
which amounts to choosing $\lambda = s $. Inserting now the required $f = 2 (U-E) + g^{ij} \d_i F \d_j F$, the variation $\delta f$ can be evaluated as, 
\begin{align}
\delta f & = 2 \delta U   - (\delta g_{kl}) g^{ki} g^{lj} \d_i F \d_j F \nn \\
& =   2 \int d x^3 \left[ - \Delta \fs \dphip + V'(\fs) \dphip  -   \d_k \fs \d_l (\dphip) ~\d^k F \d^\ell F \right]  \, ,
\end{align}
where in the first line we used that $\delta F = 0$ in the first line according to Eq.~\eqref{eq:firstEoM}. The second  line then follows via an integration by parts with respect to space. 
We can then straightforwardly use Eq.~\eqref{eq:perpphi} to find
\begin{align}
\delta f & =  2 \int d x^3 \left[ - \Delta \fs \dphip + V'(\fs) \dphip \right]  + 2 \int d x^3 \dphip (\d^i F \d^j F \d_{i}\d_{j} \fs)   \, .
\end{align}
Inserting back into Eq.~\eqref{eq:perpphi2} leads to,
\begin{equation}
\label{eq:IntegratedEoM}
\int d x^3 \int_{\lambda_i}^{\lambda_f} d \lambda  (\dphip) \left[ - \d_\lambda^2 \fs  - \Delta \fs  + V'(\fs)  +  \d^i F \d^j F \d_{i}\d_{j} \fs \right]= 0 \, ,
\end{equation}
This then gives us the full equation of motion to be satisfied by the field,
\begin{align}
\label{eq:EoM_full}
(\d^i G \d^j G - \d^i F \d^j F)  \d_i \d_j \fs   &~\equiv ~- \Delta \fs + V'(\phi) ~\left[ +k(\lambda,\ell)\d_i \fs \right] \, ,
\end{align}
where the last bracket indicates that this equation is defined up to a field and $x$ independent multiple $k$ of $~\d_i \fs$. 
This ambiguity arises because we are only interested in minimising $G$ with respect to the orthogonal perturbation, so that one can a priori use Eq.~\eqref{eq:perpphi} and Eq.~\eqref{eq:IntegratedEoM} to add an arbitrary multiple of the single derivative factor of the form  $\int d x^3  \d_i \fs  ~\dphip $. It is straightforward to check that $\delta_\perp F$ leads to same equation.

\bigskip 

The full set of equations to be solved is then
\begin{align}
\label{eq:FullSet}
(\d^i G \d^j G - \d^i F \d^j F)  \d_i \d_j \fs   &~= ~- \Delta \fs + V'(\phi) ~\left[ +k(\lambda,\ell)\d_i \fs \right] \\[0.5em]
(\d^i G \d^j G - \d^i F \d^j F) g_{ij} &~=~ 2 (U-E) \label{eq:FullSetEne} \\[0.5em]
\d^i G \d^j F ~g_{ij} &~=~ 0  \label{eq:FullSetProb} \\
\label{eq:prob}
g_{ij} &~=~ \intv dx^3 \d_i \fs \d_j \fs \, .
\end{align}

Let us conclude this section by making some important comments.  First, fixing the norm $m_{\lambda}$ amounts to fixing the parametrisation along the G-lines. Or, equivalently, partially fixing the diffeomorphism invariance of Eq.~\eqref{eq:FSE3}. At zeroth order, since $f = 2 (U-E)$, Eq.~\eqref{eq:gaugefix} is simply energy conservation in Euclidean time. Hence, in the FSE formalism, (Euclidean) time is defined as being the parametrisation respecting (Euclidean) energy  conservation.  Second, the hyperplane $\scr{H}$, consisting of the field configuration $\phi_s$ in field space, is constructed here from the solution of $\left.\frac{\delta G}{\delta \phi_\perp } \right|_\scr{H} =  0 $ along each G-line. As such, there is no guarantee that it should be everywhere smooth or well-defined. In particular, if we suppose that the G-lines are defined from $(t,\lambda=\lambda_i)$, there may be certain values of the parameter $t$ for which $\left.\frac{\delta  G}{\delta \phi_\perp } \right|_\scr{H} =  0 $ does not have a solution, and when solutions are found, smoothness of the resulting hyperplane can rigorously be ascertained only in a small neighbourhood of $\lambda_i$. Finally, it would be interesting to see if the equation set~\eqref{eq:FullSet} could be transformed into a form where the metric $g_{ij}$ would be directly one of the variables and not a derived quantity, we leave this possibility for future work.

\bigskip
At this point we are ready to approach a concrete tunneling problem. We will do that in the next section~\ref{sec:Oscillating}.
In the remainder of this section we give arguments to justify our procedure and in particular the use of the time-independent FSE.

\subsection{Dynamical tunneling}
\label{Sec:dynQM}
The formalism developed in the previous subsections is based on the use of a time-independent functional Schr\"odinger equation.
In this section we argue that the tunneling rate exponent can be obtained by studying a time-independent problem with suitable boundary conditions. In particular, extending this to the reduced QFT problem leads us in Sec.~\ref{sec:Oscillating} to introduce time-dependent boundary conditions, matching the field value and its time derivative.

Most of our arguments will be based on quantum mechanics but in some places we also briefly refer to the full quantum field theoretical situation.

\subsubsection{Time dependent states}

Let us start with a review of the relevant features of time-dependent states in a quantum theory.

Energy eigenstates are quasi-stationary in the sense that their corresponding probability distributions are
time-independent,
\begin{equation}
P(x,t)=|\psi(x,t)|^2=|\psi(x)|^2,\quad{\rm for\,\, energy\,\,eigenstate}.
\end{equation}

Tunneling can nevertheless be described by considering the particle flux on the left and the right hand side of the barrier.
Importantly in quantum mechanics this calculation can be done for any (permissible) energy, not only in for the ground state.
This already gives some justification for trying to consider the time independent Schr\"odinger equation also for an energy 
that is not the ground state energy.

Nevertheless there is no true time dependence yet.
In quantum mechanics time-dependence arises from a superposition of states with {\emph{different}} energy.
The states closest to the oscillating homogeneous fields considered in Sec.~\ref{sec:Oscillating} are coherent states.
In the case of a harmonic oscillator they are (see e.g.~\cite{schwabl,wiki}),
\begin{equation}
|\alpha(t)\rangle=\exp\left(-\frac{|\alpha|^2}{2}\right)\sum_{n}\frac{\alpha^{n}}{\sqrt{n!}}|n\rangle\ ,
\end{equation}
where
\begin{equation}
\alpha(t)=x_{\rm max}\sqrt{\frac{m\omega}{2\hbar}}\exp(-i\omega t) \ .
\end{equation}
Here $x_{\rm max}$ is the amplitude of the ``classical'' oscillation and we have chosen the phase such that the expectation value of $x$ is maximal at $t=0$.
The corresponding wave function is given by,
\begin{equation}
\label{coherent}
\psi_{\alpha}(x)=\psi_{0}(x)\exp\left(-\frac{|\alpha|^2}{2}\right)\sum_{n}\frac{\alpha^n}{2^{n/2}n!}H_{n}\left(\sqrt{\frac{m\omega}{\hbar}} x\right)\ ,
\end{equation}
where $\psi_{0}(x)$ denotes the wave function of the ground state.

For our purposes the important observation is that a coherent state is not an energy eigenstate but instead it is a superposition of energy eigenstates.
The expectation value coincides with the classical one,
\begin{equation}
\langle E\rangle=\frac{1}{2}m\omega^2 x^{2}_{\rm max}+\frac{1}{2}\hbar\omega\ .
\end{equation}
More importantly the variance is non-vanishing,
\begin{equation}
\Delta E^2=\hbar\frac{m\omega^3}{2}x^{2}_{\rm max}\approx \hbar\omega E\ ,
\end{equation}
where the approximate equality holds for large excitations.
As we would expect for the transition to classical behaviour the relative uncertainty in the energy decreases with increasing energy (or occupation number),
\begin{equation}
\frac{\Delta E}{E}=\sqrt{\frac{\omega}{E}}=\frac{\sqrt{2}}{\sqrt{m\omega/\hbar}x_{\rm max}}.
\end{equation}

This has important consequences for our field theoretical case of interest.
First of all it is straightforward to generalise the result of the quantum mechanical harmonic oscillator to the field theory 
Schr\"odinger functional,\footnote{To obtain a suitable finite volume we could, e.g., consider a three dimensional torus with periodic boundary conditions. We use $\hbar=c=1$.}
\begin{equation}
\Psi[\phi]=\Psi_{0}[\phi]\exp\left(-\frac{|\alpha|^2}{2}\right)\sum_{n}\frac{\alpha^n}{2^{n/2}n!}H_{n}\left(\sqrt{ m \scr{V}}\phi\right).
\end{equation}
In the field theoretical case we have the relations,
\begin{equation}
\alpha=\phi_{\rm max}\frac{\sqrt{m \scr{V}}}{2},
\end{equation}
and
\begin{equation}
\frac{\Delta E}{E}=\frac{\sqrt{2}}{\sqrt{m \scr{V}}\phi_{\rm max}} ,
\end{equation}
where $\phi_{\rm max}$ is the amplitude of the homogeneous field oscillations.
We are also interested in the (spatially averaged) field amplitude,
\begin{equation}
\frac{\Delta \phi}{\phi_{\rm max}}
=\sqrt{\frac{\sqrt{2}}{\sqrt{m{\mathcal{V}}}\phi_{\rm max}}}.
\end{equation}
This also goes down with the volume.
Crucially the relative uncertainty is not only suppressed with the amplitude of the oscillation, but also with the 
volume $\scr{V}$ that we consider.

Generally speaking for macroscopic volumes the relative energy (and field) variance is much smaller in quantum field theory.
While this is suggestive of using an energy eigenstate for the calculation of the tunneling rate if the dependence of the rate on the energy is not too big, some caution is required. First of all, the relevant volume for the bubble formation is only the size of the bubble, hence it is not infinite. Perhaps more importantly states with degenerate energy can have already quite different tunneling rates.

This can be easily seen from a textbook two dimensional quantum mechanical example.\footnote{In one dimension one usually does not have degenerate states.}
Let us consider a rectangular potential barrier with infinite extent in one direction.
Finding the tunneling solutions for this problem is exactly the same as the case of a one dimensional barrier. 
Putting the barrier in the $x$-direction the problem factorises,
\begin{equation}
\psi(x,y)=C\exp(ik_{y}y)\phi(x),
\end{equation}
where $\phi(x)$ is the tunneling solution in the one-dimensional problem and $C$ is a normalisation constant.
The tunneling probability is given by,
\begin{equation}
\label{eq:example}
P(k_{y})=\frac{1}{1+\frac{V^{2}_{0}}{4(E-\hbar^2 k^{2}_{y}/2m)(V_{0}-(E-\hbar^2k^{2}_{y}/2m))}\sinh^{2}\left(2a\sqrt{m}/\hbar\sqrt{2(V_{0}-(E-\hbar^2 k^{2}_{y}/2m)}\right)},
\end{equation}
where $a$ is the thickness of the barrier and $V_{0}$ its height.

For a given energy the tunneling probability therefore strongly depends on the size of $k_{y}$.
In a sense this is not surprising since only the momentum transverse to the barrier is relevant for the tunneling rate (the same is also true in the classical case).

In the field theoretical case there is an infinite number of degenerate states for a given energy. In general the tunneling probability will depend on the specific properties of the initial state. Importantly, for our initial state to be a proper ``classical'' state, we will later take any momenta orthogonal to the classical evolution to be negligible compared to the parallel one. In particular this fixes the degeneracy described above.

That said, Eq.~\eqref{eq:example} in a sense already exhibits structure that we will follow (similar to what is done in~\cite{Bowcock:1991dr}). In one direction we have classical motion and in the other, perpendicular one we have tunneling. We will therefore split the QFT problem into a direction with classical motion,
implementing the classical boundary conditions for the  field and its time derivative, and a tunneling direction. We hope that this captures the relevant features of the initial state. Nevertheless, in general we should keep the above caveat in mind.

\subsubsection{The saddle point approximation and initial conditions}
Starting from the wave function at an initial time $t_{i}$, $\psi(x,t_{i})$ we can determine the wave function at some later time $t_{f}$ by the path integral expression (we closely follow the arguments given in~\cite{Turok:2013dfa}):
\begin{equation}
\label{pathintegral}
\psi(x_{f},t_{f})=\int {\mathcal{D}}x\int dx_{i} \exp\left(iS[x(t),x_{f},t_{f},x_{i},t_{i}]/\hbar\right)\psi(x_{i},t_{i})\ .
\end{equation}
Here $S[x(t),x_{f},t_{f},x_{i},t_{i}]$ is the action for a path $x(t)$ that has initial values $x_{i}$ at $t_{i}$ and final value $x_{f}$ at $t_{f}$.

Let us now consider a situation where the initial wave function is given by the coherent state Eq.~\eqref{coherent}. For convenience we can write it as,
\begin{equation}
\psi(x,t)={\mathcal{N}}\exp\left(-\frac{m\omega}{2\hbar}\left(x-x_{cl}(t)\right)^2+ip_{cl}(t)x/\hbar+i\theta(t)\right)\ ,
\end{equation}
where $x_{cl}(t)$ and $p_{cl}(t)$ are the ``classical'' position and momentum of the position and momentum at time $t$. In the quantum mechanical setup they coincide with the expectation values of the respective quantities.

We can now employ the saddle point approximation in the variable $x_{i}$ (both $t_{i}$ and $t_{f}$ are given and fixed).
Minimising the exponent in Eq.~\eqref{pathintegral} we have,
\begin{equation}
i\frac{\partial{S}}{\partial x_{i}}-m\omega (x_{i}-x_{cl}(t_{i}))+ip_{cl}(t_{i})=0 \ .
\end{equation}
Using,
\begin{equation}
\frac{\partial S}{\partial x_{i}}=-p_{i}
\end{equation}
we therefore have,
\begin{equation}
m\omega(x_{i}-x_{cl}(t_{i}))+i(p_{i}-p_{cl}(t_{i}))=0 \ .
\end{equation}

Insisting that both $x_{i}$ and $p_{i}$ are real we obtain the initial conditions for our time-dependent problem,
\begin{equation}
x_{i} =x_{cl}(t_{i}),\qquad p_{i}=p_{cl}(t_{i})\ .
\end{equation}

Using the methods developed in Sections~\ref{sec:FSE} and~\ref{Sec:multiRed} we will implement these boundary conditions in a simple example in Sec.~\ref{sec:Oscillating}.

\subsubsection{Superposition of states and its effect on the tunneling rate}
\label{sec:QMandWKB}

As we have seen above the relative energy spread in quantum field theory is not very large, nevertheless there is an important caveat to keep in mind.
The tunneling rate does in general not only depend on the energy of the system. However, if one assumes that the potential orthogonal to the initial direction of the oscillations is very steep, the system can be assumed to be in the ground state in this direction.\footnote{Or at least very near to the ground state in the sense that excitations in these directions are ${\mathcal{O}}(\hbar)$ compared to the evolution in the ``classical'' direction. This is particularly relevant for the field theoretic case where perturbations around the classical homogeneous solution are $\hbar$-suppressed.}

We can then develop a picture (cf.~\cite{Bowcock:1991dr}) where we have classical evolution in one direction (in our case $y$) and the tunneling direction is in the other variables.

Let us first focus on creating a suitable initial state for our problem. We consider a two-dimensional quantum mechanical problem with potential $V(x,y)$.\footnote{Which can for instance correspond to the effective potential introduced earlier restricted to the hypersurface $\scr{H}$, with $U[\fs (\lambda,\ell)]$.} We will suppose that the system is initially classically evolving in the $y$ direction. As discussed above, we decompose the initial wave function as:
\begin{align}
\Psi_i = \Psi_{x,0} \Psi(y,t) \, ,
\end{align}
where $ \Psi_{x,0}$ is a ground state solution for the approximately harmonic steep potential in the $x$ direction. 

We are interested in forming a wave packet of time-independent WKB solutions around a coordinate $y_0$ and therefore define
\begin{align}
\Psi(y,t) ~\propto~ \exp \left[\frac{i}{\hbar} F(y) - \frac{i}{\hbar}  E t \right] \ .
\end{align}
Varying the energy,  the available impulsion $k$ at a given point $y_0$ can be defined through
\begin{align}
\d_y F |_{y=y_0}  ~\equiv~ k(y_0) ~=~ \sqrt{2 m (E- V_0 )}  \, ,
\label{eq:momentumfix}
\end{align}
where we have noted $V_0 = V (y_0)$ and the second equality derived from the Schr\"odinger equation at $y_0$.\footnote{Notice that we only consider right-moving positive solutions for F, but the reasoning would proceed similarly for left-moving solutions.} Reciprocally, we can label these WKB solutions by their momentum $k$ at $y_0$, using that their energy satisfies at the point $y=y_0$,
\begin{align}
E(k) = V_0 + \frac{k^2}{2 m} \, .
\end{align}

We stress that this is merely a way of labelling the one-dimensional family of the WKB solutions of the Schr\"odinger equation for the potential $V(0,y)$. Focusing on an initial oscillating state, we can form a Gaussian wave-packet of variance $\lambda$:

\begin{align}
\label{eq:IniPsi1}
\Psi(y,t) \propto \int d k \exp \left[ - \frac{(k_0-k)^2}{2 \lambda \hbar^2} + \frac{i F }{\hbar} - i \frac{ E (k) t }{\hbar} \right] \, .
\end{align}
If we expand $F$ around $y_0$ as 
\begin{align}
\label{eq:Fexpansion}
F (k,y) = F (E_0,y_0) + (y-y_0) k + \frac{k^2-k_0^2}{2 m} \d_E F|_{E_0,y_0}  \, ,
\end{align}
we obtain after integrating
\begin{align}
\Psi(y,t) \propto   \exp  \left[ - \frac{\lambda m^2}{2 (m^2 +\lambda^2 \hbar^2 t^2)} \left(   (y_0+ \frac{ k_0}{m} t ) - y \right)^2 + \frac{i \alpha }{\hbar} \right] \ ,
\end{align}
where we have translated the time parameter by the constant $\d_E F|_{E_0,y_0}$ and collected in $\alpha$ the terms contributing to the phase. We obtain a standard Gaussian wave packet of plane waves centred in momentum around $k_0$ and in position around $y_0$ with a spread controlled by $\lambda$, as constructed in a similar context in~\cite{Bowcock:1991dr}. In particular, the choice $\lambda = \scr{O} (\hbar^{-1})$ leads to wave packet localised in both position and momentum with a $ \scr{O} (\hbar)$ spread. Notice that the localisation of the crest of the wave packet could also be easily obtained by using the saddle point approximation on the exponent of~\eqref{eq:IniPsi1}, leading to,
\begin{align}
\begin{cases}
\quad   k-k_0 = 0 \\
\quad     y-y_0 = \displaystyle \frac{k}{m} t  \, ,
\end{cases}
\end{align}
and we recover the complete wave packet results.

Once $F$ and $G$ have been obtained by solving the time-independent Schr\"odinger equation, the wave function in the full system can be obtained from the original wave packets as,
\begin{align}
\Psi(x,y,t) \propto  \int dk  \exp \left[ - \frac{(k_0-k)^2}{2 \lambda \hbar^2} + \frac{i F }{\hbar} -  \frac{ G }{\hbar}- i \frac{ E (k) t }{\hbar} \right]  \, .
\end{align}
As was already pointed out in~\cite{Bowcock:1991dr} the wave packet shape itself is strongly deformed during the tunneling due to the simple fact that each part of the packet feels a different part of the barrier. 

Let us nevertheless study the dependence of the crest of the packet on the other coordinates $x$ (effectively the tunneling directions) at a fixed time $t$ using again the saddle point approximation. 
Effectively we are asking how a wave packet localized around a classical path (and therefore being maximal on it) extends into the remaining directions.

The important observation here is that $F$ is constant along the lines generated by $\vec{\nabla} G$, henceforth called G-lines (see Appendix~\ref{sec:tunQM} or~\cite{Bowcock:1991dr}),
so that 
\begin{align}
F(x,y,E) = F\big(0,Y_0(x,y,E)\big)   \, ,
\end{align}
where $Y_0$ is the inverse function along G-lines defined such that given a point $\vec{x}_1$, $Y_0(\vec{x}_1,E) = y_0$ along the G-line starting at $(0,y_0)$. Applying the previous decomposition~\eqref{eq:Fexpansion} of $F(0,y)$ in this case leads to the deformed expansion
\begin{align}
F(\vec{x}_1+\vec{\delta x},E) = F\left(0,y_0,E_0\right) + \vec{\delta x} \cdot \vec{\nabla} Y_0 |_{\vec{x}_1}  k  + \frac{k^2-k_0^2}{2 m} \left(\d_E F|_{E_0,y_0}+ k \d_E Y_0  |_{\vec{x}_1} \right) \, ,
\end{align}
and finally, using the saddle point approximation, and again translating the time coordinate to absorb the constant  $\d_E F|_{E_0,y_0}$, we obtain
\begin{align}
\begin{cases}
\quad   k-k_0 = \displaystyle \lambda \hbar \frac{k}{m} \d_E G \\
\quad     \vec{\delta x} \cdot \vec{\nabla} Y_0 |_{x_1,y_1} = \displaystyle \frac{\d_E Y_0 }{2 m} (3k^2-k_0^2 ) \, .
\end{cases}
\end{align}

In particular, we see that the crest of the wave packet naturally extends orthogonally to $ \vec{\nabla} Y_0$ -- and therefore along the G-lines--, but can be moved away by corrections stemming from $\d_E G$ and $\d_E Y_0$.\footnote{In the QFT case the energy derivative are naturally suppressed by the volume $\scr{V}$ compared to the coordinate ones $\d_E \propto 1/\scr{V}$.} 

Overall we conclude that as a first approximation, we can estimate the time-dependent tunneling rate for a classical oscillating state $y(t)$ of energy $E$ by determining $G(x,y)$ for a WKB state of energy $E$ and then following the integral line of $G$ starting at $y(t)$.

When moving into the full-fledged quantum field theory, this conclusion could be modified in two ways. First, the field support on which we will project the problem onto a quantum mechanical will depend on the energy, implying that during tunneling, the wave packet will also spread in field space. However, since the initial wave function is only one-dimensional, the field support will be independent of the energy (as one will be always able to reparametrise time to ensure it), and depends only on the potential. This implies that the initial wave packet can be formed on a one-dimensional support as we did for the quantum mechanical problem. We will neglect the subsequent spreading effect in the following.  Second, the mass term $m$ defined above is now a function of the metric $g_{ij}$ on the field hyperplane $\scr{H}$, so that one has $m(x,y)$. In practice, we then define $m=m(0,y_0)$, ensuring that the initial wave packet is properly defined. The subsequent effect of the variation of $m$ is then included in $Y_0$ since the shape of the G-lines depends on the metric.

\section{Tunneling in an oscillating background}
\label{sec:Oscillating}

As an application of the formalism described above, we evaluate the  time-dependent decay rate of an initially homogeneous field configuration oscillating around a false vacuum to a deeper one. The two-dimensional hyperplane $\scr{H}$ will be constructed from all the MPEP at a given time of the oscillation.

\subsection{The membrane action}

We consider a setup similar to the one introduced in~\cite{Darme:2017wvu}. Namely, we use an asymmetric double well potential of the form 
\begin{align}
\label{eq:potential}
V = \frac{g c^4}{4} (\phi^2/c^2 - 1)^2 - B (\phi + c) \, ,
\end{align}
where $g,c$ and $B$ are positive constants. We define the inverse ``thickness'' of the wall by
\begin{align*}
\mu \equiv \sqrt{2 g c^2} \, ,
\end{align*}
and assume the ``thin-wall'' hierarchy,
\begin{align}
\label{eq:thinwall}
B c  \ll \mu^2 c^2 \, .
\end{align}
It will be useful to introduce the thin-wall parameter $\alpha$ defined by,
\begin{align}
\label{eq:alpha}
\alpha \equiv \frac{B c}{\mu^2 c^2} \ll 1 \, ,
\end{align}
such that the radius of the vacuum-to-vacuum bubble $R_0$ is given by,
\begin{align}
R_0 = \frac{1}{\mu \alpha} \ .
\end{align}
During the initial oscillating phase, the field undergoes a classical evolution with its energy density conserved. The energy density of an oscillating scalar field is given by
\begin{align}
\label{eq:ED}
e = \frac{1}{2} (\d_t\phi)^2 + V(\phi)
\end{align} 
and remains constant during the oscillations since kinetic energy is simply transferred to potential energy. Hence, provided that we can neglect the variation in tension (small oscillations) the initial radius of the classical solution corresponding to a true vacuum bubble will not depend on the time at which tunneling occurs. According to the discussion of Sec~\ref{Sec:dynQM}, we will focus on solving the FSE in the WKB regime, assuming a stationary solution with energies very close to the classical energy of an oscillating initial state.

In the thin-wall limit, we can neglect the details of the potential shape and \linebreak parametrise the whole evolution of the system as a function the membrane tension $\sigma_0$ along the wall $\scr{W}$, the difference between the energy density in the membrane and outside of it $\vep$, the bubble radius $R$ (we suppose a spherically symmetric bubble), and the value of the field outside of the bubble (assumed to be unperturbed). As noticed in~\cite{Darme:2017wvu}, the field oscillations within the bubble are typically suppressed during the bubble nucleation process, we will therefore assume we can neglect them in the following. Furthermore we will consider that the field outside the bubble is not perturbed by the bubble nucleation and given by small harmonic oscillations around the false vacuum,
\begin{align}
\label{eq:fieldout}
\phi_{out} = c (-1 + q_f \cos \mu T ) \, ,
\end{align}
where we use upper-case $T$ to emphasize the fact that this time parameter is used to describe the field configuration outside the bubble. The energy of such a classical initial state over the control radius $\Lambda$ is simply
\begin{align}
E = \frac{4}{3} \pi \Lambda^3 e \, .
\end{align}

Crucially, the dominant expansion parameter in this scenario is not $q_f$, but rather the ratio $q_f^2/\alpha$. As argued in~\cite{Darme:2017wvu} this expansion parameter roughly compares the size of the oscillations with the thin wall approximation and ensures that the subsequent evolution of the bubble is not strongly modified by the surrounding oscillating field. Similarly, we will see that when considering the variation of tunneling rate, the expansion naturally orders around  $q_f^2/\alpha$. In the thin-wall regime, the direct consequence of requiring  
\begin{align*}
\frac{q_f^2}{2 \alpha} \ll 1 \ ,
\end{align*}
is that all changes of the wall tension derived from the oscillation of the external field, proportional to $q_f^2$ are negligible at first order. The radius of the bubble nucleated at the extremum of the oscillation $R_e$ has been determined in~\cite{Darme:2017wvu} using the standard Coleman instanton approach. At first order in  $q_f^2/\alpha$ it reads,
\begin{align}
R_e = \frac{R_0}{1+\frac{q_f^2}{4 \alpha}} \ . 
\end{align}

Based on these assumption, it is possible to reduce the action for the scalar field $\phi$ to the simplified form~\cite{Darme:2017wvu},
\begin{align}
\label{eq:Lmembrane}
S = \int dT\left[-4 \pi \sigma_0 R^2 \sqrt{1-\dot{R}^2} + \frac{4}{3} \pi p R^3  + \frac{4}{3} \pi p_{\out} \Lambda^3  \right] \, ,
\end{align}
where the pressure  is defined by
\begin{align}
\label{eq:pressure}
p  \equiv p_{\ins} - p_{\out} \equiv \left(\frac{1}{2} \dot{\varphi}_{\ins}^2 - V(c+\phiin)\right) - \left(\frac{1}{2} \dot{\varphi}_{\out}^2 - V(-c+\phiout) \right) \, . 
\end{align}
In particular, for the potential~\eqref{eq:potential} introduced earlier, the tension can be written as,
\begin{align}
\label{eq:tension2}
\sigma_0 =  \frac{2 \mu }{3} \left[ c^2 + \scr{O} \left(  \phiin^2,\phiout^2 \right) \right]\, .
\end{align}
In this form, and as it was noticed in~\cite{KeskiVakkuri:1996gn} the action strongly resembled the one of particle pair creation in a time-dependent electric field, with the distance between electron and positron replacing the radius $r$ of the bubble. The latter can be solved using the standard instantonic method by promoting time to a complex parameter and solving the corresponding equations of motion. However, these two scenarios differ in a crucial way: for the oscillating tunneling case, the time dependence is given by the initial state dynamics, which is mainly now hidden in the expression for the pressure in Eq.~\eqref{eq:Lmembrane}. In contrast for the pair creation case, it is the potential which depends on time due to the presence of the background electric field. In that sense, the oscillating field tunneling resembles more closely the QM case with a initially oscillating particle~\cite{Turok:2013dfa} and the pair creation process, the tunneling in a time-dependent potential~\cite{Widrow:1991xu,Simon:2009nb}. One issue with the membrane form of the action~\eqref{eq:Lmembrane} is that it does not properly describe the initial configuration of the field through the variable $R$, so that the setup is markedly different from the one described in~\cite{Turok:2013dfa}. In particular, it is not clear that complexifying the time parameter will properly describe the bubble tunneling.

The expression~\eqref{eq:tension2} corresponds to the tension for the vacuum-to-vacuum, for which one can easily find an approximate form for the bubble's field profile. Indeed, following the standard treatment from~\cite{Callan:1977pt,Coleman:1980aw} and assuming that the solution during tunneling is $O(4)$-symmetric, the field only depends on $\xi  \equiv \sqrt{\lambda^2+|\vec{x}|^2}  \equiv \sqrt{\lambda^2+\rho^2}$ and Euclidean equation of motion for the field profile $\phi$ is,
\begin{align}
\label{eq:colemanEOM}
\frac{\d^2 \phi}{\d \xi ^2} +  \frac{3}{\xi } \frac{\d \phi}{\d \xi } = V'(\phi) \ .
\end{align}
The first ``viscous'' derivative term in~\eqref{eq:colemanEOM} is neglected in the ``thin-wall approximation'' around the bubble wall, leading to the  usual solution
\begin{align}
\label{eq:phiColeman} 
\phi_0(\xi ) = -c \tanh \left(\frac{\mu}{2} (\xi -R_0) \right) \ ,
\end{align}
where $R_0$ is the final radius of the bubble defined above. An important comment is that~\eqref{eq:phiColeman} can be further approximated in the vicinity of the bubble by noting that 
\begin{align}
\xi -R_0 \simeq \frac{\rho^2 - R^2}{2 R_0} \ ,
\end{align}
where $R = \sqrt{R_0^2 - \lambda^2}$ is the radius of the bubble during tunneling (namely between $\lambda = -R_0$ and $\lambda = 0$ in the chosen parametrisation). Using this parametrisation, we can put aside the imaginary time parameter $\lambda$ altogether and describe the nucleating bubble directly for its 3D field profile along with its radius. Notice also that using this approximation, the viscous terms is suppressed by a thin-wall parameter $\alpha$ compared to the second-derivative one, validating the consistency of our approximation.

In the following section, we will use the approach described in the previous section based on the FSE instead of the instantonic method. Our final result then resembles more closely the results from~\cite{Turok:2013dfa} then the one for tunneling in a time-dependent potential in that the tunneling rate exponent will not have an exponential increase with the oscillations.

\subsection{The FSE approach}

The first step is to use the membrane approximation to simplify the equations of motion for the field and absorb the space-dependence. Since we neglect the variation in tensions and following the discussion in the previous sections. 
We will assume that along the wall the solution can be parametrised with $r$ corresponding to $\xi - R_0$ in Eq.~\eqref{eq:phiColeman}, and being given by
\begin{align}
r ~\equiv~  \frac{\rho^2 - R(\lambda,\ell)^2}{R_f(\ell)} \ .
\end{align}
Here $R$ is the radius of the bubble and $R_f$ is the final radius after tunneling, assumed to be a function of $\ell$.

The choice mostly amounts to fixing the dependence on $\rho$ in our bubble profile, and assuming that this profile will be deformed during the tunneling.
The radius $R$ of the bubble depends on the variables parametrising the two-dimensional surface in field space which we use for our tunneling process. Using this form amounts to neglecting all the derivatives which are parallel to the wall and taking only the variation perpendicular to it into account. Furthermore, we continue to use the thin-wall approximation so that we can neglect first derivative terms in $\rho$ compared to the second order derivative contributions. 

We will use a short-hand notation 
\begin{equation}
\dlam \equiv \d_i G \d_i \quad{\rm and}\quad \dell \equiv \d_i F \d_i \ .    
\end{equation}
The equation for the field from~\eqref{eq:FullSet} then reads,
\begin{align}
\left[ (\dlam r)^2 - (\dell r)^2 + (\d_\rho r)^2 \right] \d_r^2 \phi  + f (\ell, \lambda) \d_r \phi = V'(\phi)\, ,
\end{align}
where we have included the free function $f$ to make explicit the freedom present in Eq.~\eqref{eq:FullSet} to add a multiple of $\d_i \phi$. Multiplying by $\d_r \phi$, we see that a particularly attractive choice for $f$ is
\begin{align}
f(\ell,\lambda) = \left. \frac{1}{2}\d_r \left[  (\dlam r)^2 - (\dell r)^2 + (\d_\rho r)^2 \right] \right|_{\rho \sim R(\ell,\lambda)}  \, .
\end{align}
Using this and integrating over $r$, we can find a simple relation which holds in the vicinity of the wall,
\begin{align}
\left[ (\dlam r)^2 - (\dell r)^2 + (\d_\rho r)^2 \right] (\d_r \phi)^2 = 2 V(\phi) \, .
\end{align}
In particular, we can express the wall tension as
\begin{align}
\sigma &\equiv \int_{\phi_{out}}^{\phi^{in}} d \phi \sqrt{2 V(\phi)}  = \int_{\scr{W}} d \rho ~(\d_r \phi)^2   (\d_\rho r)^2 \sqrt{ 1 + \frac{(\dlam r)^2 - (\dell r)^2}{(\d_\rho r)^2}} \, ,
\end{align}
and more importantly, absorb the space-dependence of the solution along the wall using our definition of $r$ to obtain
\begin{align}
\label{eq:tens_simp}
\int_{\scr{W}} d \rho ~(\d_\rho \phi)^2  =  \frac{\sigma_0 }{ \sqrt{ 1 + (\dlam R)^2 - (\dell R)^2}} \ ,
\end{align}
where in replacing $\sigma$ by $\sigma_0$, we used the fact that the variation of the tension is proportional to  $q_f^2$ and therefore negligible at first order in $q_f^2 / 2 \alpha$.
Let us now turn to the FSE equations in Eq.~\eqref{eq:FullSet}-\eqref{eq:prob}. The second step is to use the previous result to expand explicitly the effective potential part. We find
\begin{align}
2(U - E) &= (4 \pi \sigma_0) R ^2 \left( -2 \frac{R}{R_e} + \frac{R^3 - \Lambda^3}{R_0 R^2} \frac{q_f^2}{2\alpha} \sin^2 \mu T   \right) + 4\pi^2 R^2\int_{\scr{W}} d \rho \left[ (\d_\rho \phi)^2 + 2V \right] \\
& = (4 \pi \sigma_0 R^2) \left( -\frac{R}{R_e} + \frac{2+(\dlam R)^2 -(\dell R)^2}{\sqrt{1+(\dlam R)^2 -(\dell R)^2}} - m_T\right).
\end{align}

Finally, we can recast the energy conservation, Eq.~\eqref{eq:FullSetEne}, and probability conservation,  Eq.~\eqref{eq:FullSetProb}, equations in two different ways. The first, more general possibility would be to use the G-lines approach outlined earlier in Sec.~\ref{Sec:multiRed} and using the definition of the parameter $\lambda$ to express Eq.~\eqref{eq:FullSetEne} and Eq.~\eqref{eq:FullSetProb} as,
\begin{align}
\d_\lambda G &=  2(U - E) +  \d^i F \d_i F \\
\d_\lambda F &= 0.
\end{align}
The second equation implies that $F$ is constant along a G-line and thus equal to its initial value at a time $T_0 ( T , R)$ where $T_0$ is the function giving the initial time corresponding to the G-line passing by the point $(T,R)$.  

Notice that the term $\d^i F \d_i F$ includes a contribution from the induced metric on the hypersurface $\scr{H}$. A convenient parametrisation of $\scr{H}$ is to label the field profiles using $(T,R)$. The parameter $T$ fixes the field outside of the bubble as in Eq.~\eqref{eq:fieldout}. It can be matched to the time parameter of the time-dependent FSE using wave-packet procedure described in Sec.~\ref{sec:QMandWKB}. We first introduce the metric elements for the $R$ and $T$ parameters by
\begin{align}
m_R \equiv \intv d^3 x (\d_R \phi)^2 = \frac{(4 \pi \sigma_0 R^2)}{\sqrt{1 - (\dlam R)^2 + (\dell R)^2}} \\
m_T \equiv \intv d^3 x (\d_T \phi)^2 = (4 \pi \sigma_0 ) \frac{\Lambda^3-R^3}{R_0} \frac{q_f^2}{2 \alpha} \sin^2 \mu T \ .
\end{align}
The first equation can simply be obtained by plugging in our ansatz for the bubble wall. For the second equation we observe that the main contribution $\sim q^2_{f}/\alpha$ is simply the time derivative of the field outside the bubble but inside the control volume. The time derivative of the bubble solution integrated over the wall region is suppressed by a thin-wall factor $\alpha$ and therefore negligible in our approximation.
The off-diagonal metric components vanish as a consequence of neglecting the variation in tension. 
Following the method described in Sec.~\ref{Sec:multiRed} we can then search for the parameter equation for the G-line $(T(\lambda),R(\lambda))$.
Once all G-lines have been found, one can then extract $T_0 (T,R)$ and iterate the process until convergence. The final tunneling rate is then obtain by integrating along G-lines
\begin{align}
G = \int_{\lambda_i}^{\lambda_f} d\lambda\, \mlam \  = \int_{\lambda_i}^{\lambda_f} d\lambda \left( (\dlam R)^2 m_R + (\dlam t)^2 m_T\right) \ .
\end{align}

\bigskip

However, given that the system at hand has been substantially simplified already there is a more direct approach.
We can directly solve for $G$ and $F$ by observing that we have
\begin{align}
\dlam T & = \frac{\d_T G}{m_T} &&& \dlam R = \frac{\d_R G}{m_R} \\
\dell T & = \frac{\d_T F}{m_T} &&& \dell R = \frac{\d_R F}{m_R}  \ .
\end{align}
Note that the equations in $R$ are non-trivial since $m_R$ has some $\dlam R$ and $\dell R $ dependence. Solving them leads to an expression for $m_R$ as function of $F$ and $G$ as
\begin{align}
m_R = \sqrt{ (4 \pi \sigma_0)^2 R^4 + (\d_R F)^2 - (\d_R G)^2} \ .
\end{align}

We can now express directly Eq.~\eqref{eq:FullSet} as an equation on $F$ and $G$ as function of $T$ and $R$. After some algebra we have,
\begin{align}
\label{eq:fullFG1}
\d_T F \d_T G &= -\frac{m_T}{m_R} \d_R F \d_R G \\
\label{eq:fullFG2}
(\d_R G)^2 - (\d_R F)^2 &= (4 \pi \sigma_0)^2 R^4 \left[ 1 -\left( \frac{R}{R_e} - \frac{(\d_T F)^2 - (\d_T G)^2-m_T^2}{8 \pi \sigma_0 R^2 m_T} \right)^2 \right].
\end{align}
The boundary conditions for Eq.~\eqref{eq:fullFG1} are as follows,
\begin{equation}
G=0,\qquad {\rm and}\qquad \d_T F= (4 \pi \sigma_0 ) \frac{\Lambda^3}{R_0} \frac{q_f^2}{2 \alpha} \sin^2 \mu T  
\qquad{\rm at}\quad R=0,
\end{equation} 
where $\d_T F$ is given by the WKB momentum. An important comment is that this boundary condition on $F$ represents the initial classical evolution before the tunneling, and thus implements the time-dependence of our initial state in practice. In order to make the correspondence with Sec.~\ref{sec:QMandWKB} more explicit, notice that this initial condition corresponds to Eq.~\eqref{eq:momentumfix}, with the parameter $T$ corresponding to $y$ of the quantum mechanical problems. In other words, the momentum of the WKB solution is fixed to be the one of the classical solution at this point. The only additional difficulty in our case is that $T$ now refers to the best parameter choice to describe the field configuration of the classically oscillating state. The time-dependence corresponding to the parameter of the time-dependent FSE has been absorbed by forming a carefully chosen wave packet, as described in Sec.~\ref{sec:QMandWKB}.

Let us conclude this section by noticing that if we focus on the extremum tunneling case when $T=0$ the set of equations can be drastically simplified. In particular, $\d_T F = \d_T G = \d_R F = m_T = 0$ leads to a trivial solution $G_e(R)$ satisfying
\begin{align}
(\d_R G)^2  &= (4 \pi \sigma_0)^2 R^4 \left[ 1 -\left( \frac{R}{R_e} \right)^2 \right]  \ ,
\end{align}
which is fully compatible with our initial condition above and leads at the bubble nucleation to the result found in~\cite{Darme:2017wvu},
\begin{align}
G(R_e) = \frac{\pi}{4} \sigma_0 R_e^3 \ . 
\end{align}
At $T \neq 0$, $\d_T F \neq m_T \neq 0$ so that the above simplification does not occur. In the next section we will use a perturbative approach to find analytical and numerical results. 

\subsection{Perturbative expansion and numerics}

When considering the problem at first-order in $\frac{q_f^2}{2 \alpha}$ only, as was done in~\cite{Darme:2017wvu}, there are in fact only three variables we need to consider: the final radius $R_f$, expected to be a function of $T$, then $F$ and $G$. At zeroth-order, $F$ and $G$ are given by,
\begin{align}
\d_T F^0 = m_T  (R=0)\equiv (4 \pi \sigma_0 ) \frac{\Lambda^3}{R_0} \frac{q_f^2}{2 \alpha} \sin^2 \mu T \nn  \\
\d_R G^0 = (4 \pi \sigma_0) R^2 \sqrt{1 - (R/R_f)^2} \ ,
\end{align}
a numerically important point is that we do not use the vacuum radius in the definition of the $G^0$, this avoids the appearance of imaginary contributions later on and will be absorbed in the first order correction.

The final radius is determined by setting the right-hand side of Eq.~\eqref{eq:fullFG2} to zero and expanding $R_f = R_e + \delta^0_r$. At first order in $\delta^0_r$ one obtains,\footnote{We always use $\Lambda>R_{0}$.}
\begin{align}
\label{eq:corrRad}
\delta^0_r =	R_f - R_e = R_0 \frac{R_0^3 - 2 \Lambda^3}{R_0^3 -  \Lambda^3} \frac{q_f^2}{4 \alpha} \sin^2 \mu T \ .
\end{align}
In the limit of a large control volume, we find 
\begin{align}
R_f = R_0 \left( 1+ \frac{q_f^2}{4 \alpha} \cos 2 \mu T\right)^{-1} \, .
\end{align}
This corresponds to the radius obtained in~\cite{Darme:2017wvu} by considering that the oscillations were completely frozen during the nucleation of the bubble. When reducing the control volume closer to $R_0$ our first order approximation breaks down. 

\begin{figure}[!ht]
	\begin{center}
		\includegraphics[width=0.6\textwidth]{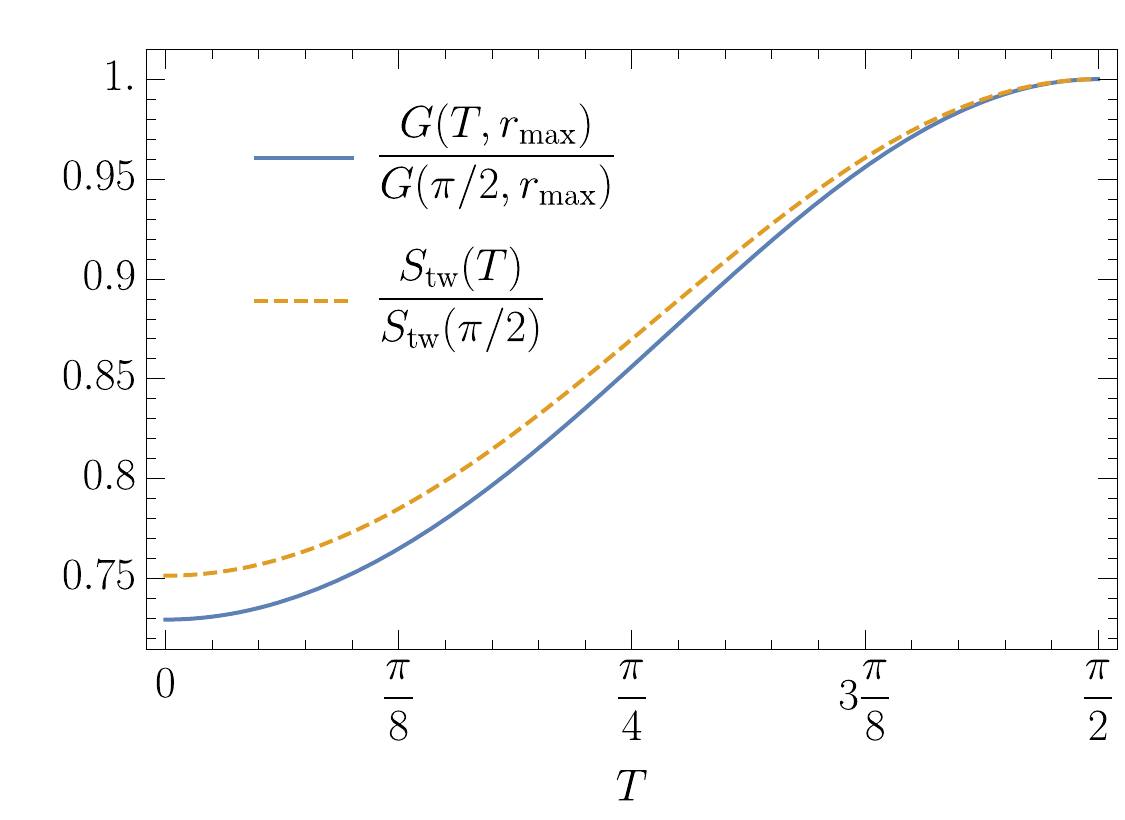}
		\caption{Comparison between time dependence of the tunneling exponent $G$ and the corresponding action $S$ obtained in~\cite{Darme:2017wvu}. The values of parameters we used in this example are $q_f^2/2\alpha=1/10$,  $4\pi\sigma_0=2$ and $R_0=1$ with $\Lambda=10$.}
		\label{fig:actionplot}
	\end{center}
\end{figure}

Replacing in the definition of $G^0$, it now simple to linearise Eq.~\eqref{eq:fullFG1} and Eq.~\eqref{eq:fullFG2} in order to fully solve the system. All the results in the following rely on the previous approach to numerically solve for $F$ and $G$. Writing then $G \equiv G_0 + \frac{q_f^2}{2 \alpha} g$ and  $F \equiv F_0 + \frac{q_f^2}{2 \alpha} f$, the system~\eqref{eq:fullFG1}-\eqref{eq:fullFG2} becomes
\begin{align}
\label{eq:firstordersyst}
\d_r f &= - R \frac{\d_T G_0 + \d_T g}{\sqrt{R_f^2-R^2}} \\
\d_r g &= \frac{R_f}{\sqrt{R_f^2-R^2}} \left[ (4 \pi \sigma_0) \left( 2 \frac{R^2}{R_e} (\delta_r - \delta^0_r \frac{R^2}{R_e^2})-\delta_r^2\right) \right.\\ 
& \left. \qquad  \qquad + \d_T f \frac{R}{R_e}\frac{\Lambda^3 }{\Lambda^3 - R^3}  \left(1 - \frac{\delta_r R_e}{R^2}\right) \right] \nn \ ,
\end{align}
where we have used the shorthand notation $\delta_r$ following the structure of Eq.~\eqref{eq:corrRad} by,
\begin{align}
\delta_r = \frac{R^2}{R_0} \frac{R_0^3 - 2 \Lambda^3}{R_0^3 -  \Lambda^3} \frac{q_f^2}{4 \alpha} \sin^2 \mu T \ .
\end{align}

This system is readily solved numerically. The value of $G$ we find at the maximal radius $r_{\rm max}$ describes the final tunneling exponent and we can directly compare it with the action $S$ we obtained numerically in~\cite{Darme:2017wvu} as shown in figure~\ref{fig:actionplot} for a control volume $\Lambda = 2 R_0$. When the control volume is significantly larger than $R_0$, the dependence is very similar and confirm the results obtained before as well as the simplifications necessary to obtain them. Indeed, this can be also seen for instance from the definition of $\delta_r^0$ which converge rapidly to a constant value at large $\Lambda$.

\begin{figure}[t]
	\centering
	\subfloat[]{%
		\label{fig:Lambdaplot}%
		\includegraphics[width=0.48\textwidth]{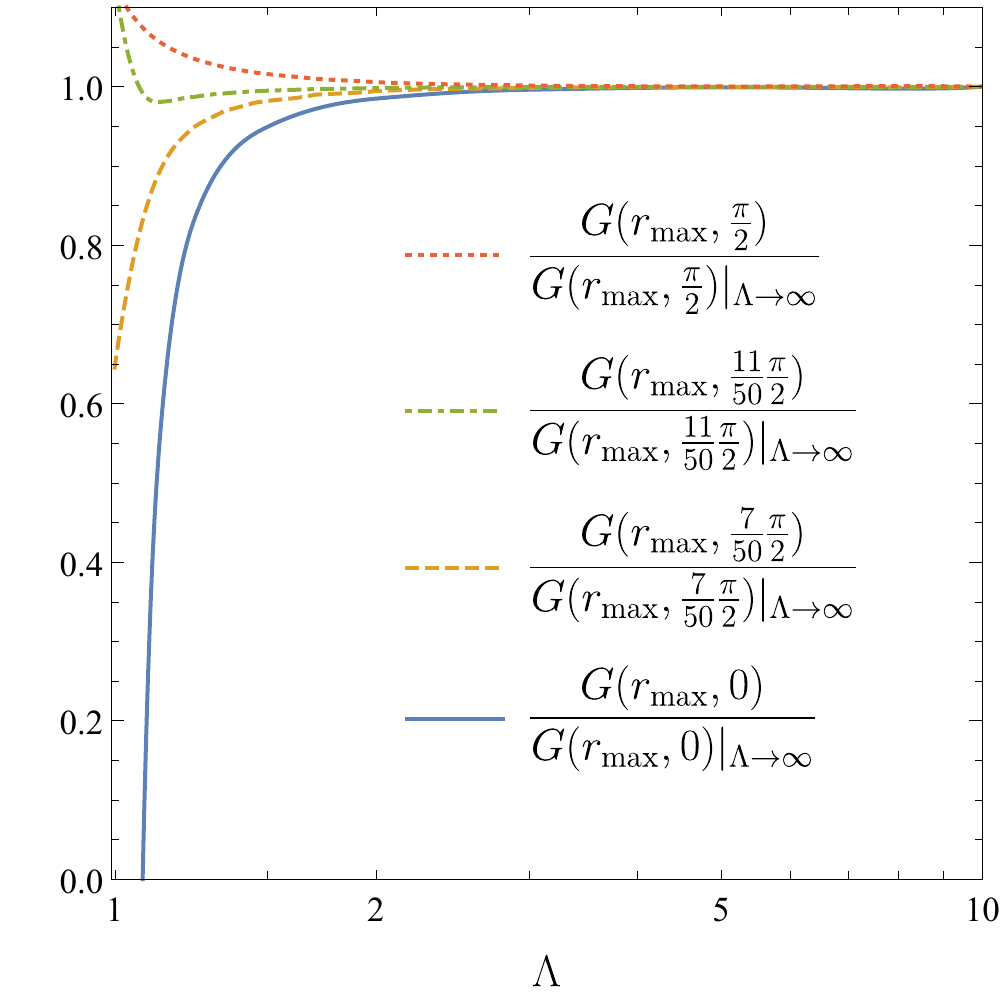}
	}%
	\subfloat[]{%
		\label{fig:Glines}%
		\includegraphics[width=0.48\textwidth]{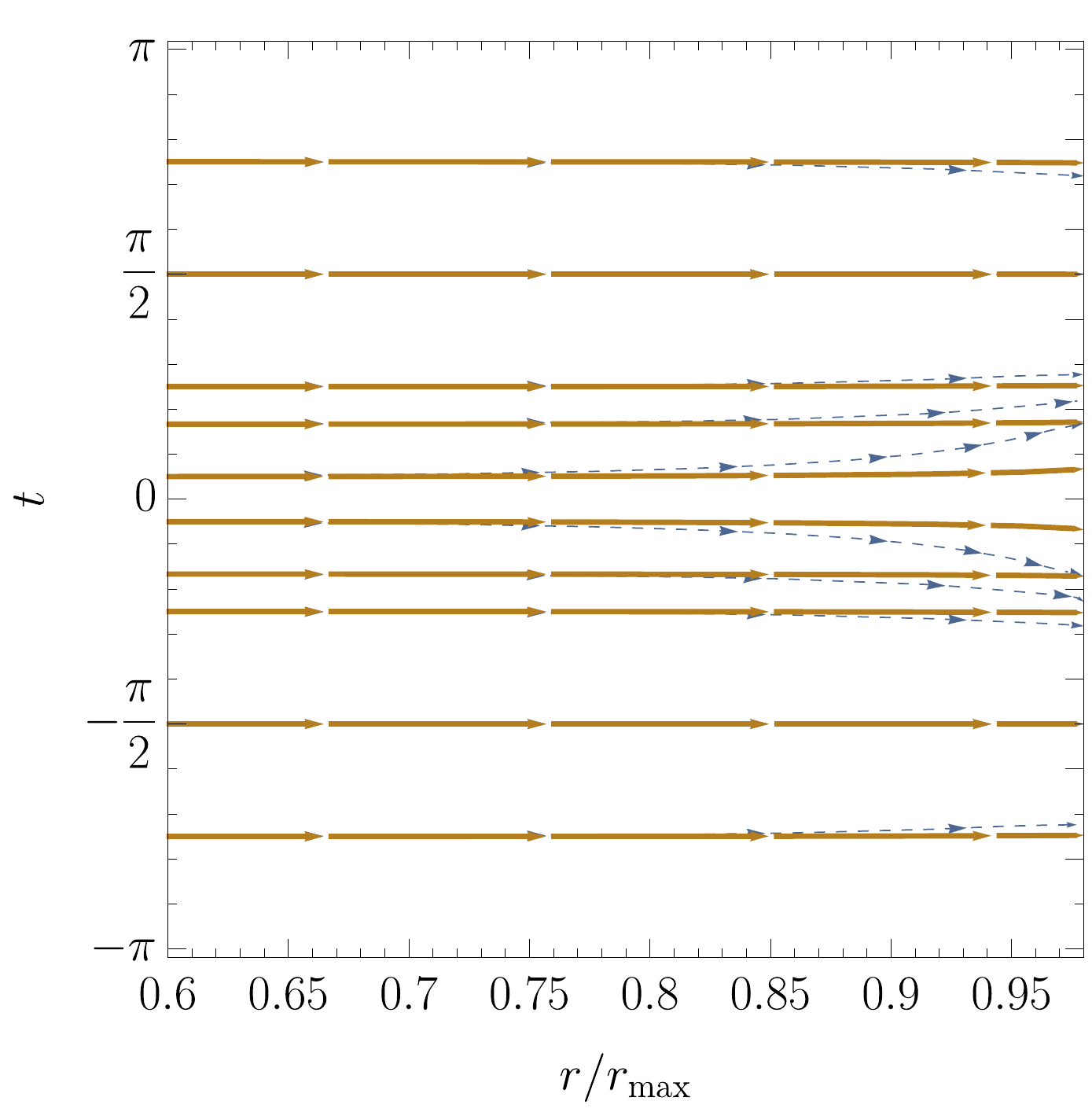}
	}%
	
	\caption{ (a) Sensitivity of the tunneling exponent to the artificial control volume parameter $\Lambda$. (b) G-lines in the R/t plane with $\Lambda=1.4$ for blue dotted lines and $\Lambda=3$ for yellow dashed lines. The values of parameters we used in both panels are $q_f^2/2\alpha=1/10$,  $4\pi\sigma_0=2$ and $R_0=1$}  
\end{figure}

We investigate in more details the sensitivity of our results to the control volume $\Lambda$ in Figure~\ref{fig:Lambdaplot}. The dependence of the final tunneling exponent on this parameter converges to the correct value very quickly as $\Lambda$ grows and is very insensitive to its precise value unless the control volume is not much bigger than the maximal bubble radius that is $\Lambda \approx r_{max}$. Finally, we show in Figure~\ref{fig:Glines} the G-lines, which as was seen in Sec.~\ref{sec:QMandWKB} represent the preferred tunneling paths at a given time.

Notice that while Figure~\ref{fig:actionplot} shows the action for $t$ between $0$ and $\pi/2$, the rest of the evolution is completely symmetrical. This is a consequence of our choice of focusing on corrections of order $q_f^2/\alpha$, and therefore neglecting variations of the bubble tension which  are typically of order $q_f^2$. Indeed, Eqs~\eqref{eq:fullFG1} and \eqref{eq:fullFG2} are fully periodic with period $\pi$ and symmetric under $T \Rightarrow \pi - T$, reflecting the symmetries of our potential. Contrary to the quantum mechanical case, having an initial state on the other side of the false vacuum merely implies that the bubble solution will be deformed to accommodate the initial value.\footnote{As an example, after expanding the equation of motion at first order around the vacuum-to-vacuum,~\cite{KeskiVakkuri:1996gn} found an approximation for the relevant bubble profiles. Note however that these profiles are classical solutions and do not per-se describe a tunneling event.}

\section{Conclusions}\label{sec:conclusions}

In this work we have presented a prescription for finding the leading contribution to the tunneling rate between an initial state dynamically evolving around an unstable vacuum and the true vacuum configuration in the framework of QFT. Based on the functional Schr\"odinger equation, we have shown how the problem can be reduced into finding a ``most probable escape hyperplane'' and solving a time-independent quantum mechanical problem on this plane. Furthermore, while the initial state is dynamical, the final tunneling rate in the quasi-classical limit can be obtained from energy eigenstate, by approximating the initial configuration from a suitable wave packet of WKB solutions.  Interestingly, we find that the quantum mechanical ``masses'' are then defined from the field configurations which mediate the tunneling. In particular they can be replaced by defining a metric on the most probable escape hyperplane. 

While noticeably simpler than the full FSE, our final set of equations, Eq.~\eqref{eq:FullSet} remains challenging to solve in full generality. Indeed, it requires solving simultaneously the field equation fixing the hyperplane as well as an integro-differential equation an this plane. Focusing on the case of tunneling from an initial oscillating state, we used the thin-wall approach to tunneling where the equation of motion for the field simplifies drastically, which allowed us to estimate at first order the tunneling rate. We have further numerically estimated this rate, hence complementing and confirming our previous results from~\cite{Darme:2017wvu} specifically devoted to this case. An appealing side-aspect of the formalism presented in this work is moreover that it could be used to describe two successive tunneling events in QFT (as argued in Appendix~\ref{sec:Twotunneling}), where the standard instanton method cannot be directly applied to describe the second bubble nucleation, leading to several controversial claims of possible resonant tunneling in this setup~\cite{henry_tye_new_2006,tye_resonant_2009}, later contested by~\cite{copeland_no_2008,Cardella:2008ww}.

While this work focuses on the direct effects of the dynamic of the initial state on quantum tunneling, the presence of parametric resonance phenomena~\cite{Kofman:1994rk,Shtanov:1994ce,Kofman:1997yn,Berges:2002cz,Berges:2002cz} (see also, e.g., the recent work~\cite{Olle:2019kbo}) which can transfer directly energy from an oscillating fields to fuel the growth of perturbations will likely also have a strong impact on tunneling by creating seeds for subsequent bubbles to nucleate. It is nonetheless important to point out that the formalism described above applies already during the first oscillations and does not require several field oscillations to build up fluctuations.

Several theoretical aspects of our calculations would nevertheless deserve a deeper look. In particular, our equations are properly defined only in a given control volume. While this was also the case for the standard vacuum-to-vacuum case, the volume terms could always be factored out. In our 2D case, the metric in the ``time-direction'' of the hyperplane depends directly on the size of the control volume. Albeit this dependence cancels out in the large volume limit, our approximation breaks down when considering a control volume close to the final bubble radius. We believe this issue could be tied with the problem of decoherence as the control volume can be seen as the typical volume on which we are able to maintain quantum coherence for long-enough to allow the tunneling process to happen. Finally, tunneling in quantum field theory has been historically tackled both through the FSE formalism and a path integral formulation, since we present in this work a study using the former, it would be interesting to check our results using with the latter.

\bigskip
\noindent \textbf{Acknowledgments}
\medskip

We thank G. Bossard, B.~Garbrecht, E.~Keski-Vakkuri and P.~Kraus for helpful discussions. LD is supported in part by the National Science Center (NCN) research grant No.~2015-18-A-ST2-00748. The work of ML was supported by the UK STFC Grant ST/P000258/1 and by the Polish National Science Center grant UMO-2018/31/D/ST2/02048.

\newpage
\appendix

\section*{Appendices}
\addcontentsline{toc}{section}{Appendix}

\section{WKB approximation}
\label{sec:tunQM}
Since our FSE approach to dynamical tunneling in QFT is based on a reduction of the problem to a simpler quantum mechanical one, it will be instructive to review shortly the basics of tunneling in QM with the Wentzel-Kramers-Brillouin (WKB) formalism, further expanding them to the case of multi-dimensional tunneling described by~\cite{Bowcock:1991dr} whom we closely follow.

Following the WKB intuition, we look for a solution of $\Psi$ of the form
\begin{align}
\Psi \propto \exp (\frac{i S}{ \hbar}) \ ,
\end{align}
where
\begin{align*}
S =  F + i G \ . 
\end{align*}
Replacing in the time-independent Schr\"odinger we obtain the system,
\begin{align}
\begin{cases}
\hbar G^{\prime \prime} + (F^{\prime}{}^2 - G^{\prime}{}^2) = 2m (E-V(\lambda) ) \\
2 F^{\prime} G^{\prime} = G^{\prime \prime} \hbar \ ,
\end{cases} 
\end{align}
where we denote derivatives with respect to $\lambda$ by a prime. It can be solved in the semi-classical limit by considering the decomposition 
\begin{align*}
F &=  \sum_{n=0}^\infty \hbar^n F_n \\
G &=  \sum_{n=0}^\infty \hbar^n G_n \ .
\end{align*}
At first order in $\hbar$ and in the classically accessible region the first non-zero coefficients are
\begin{align*}
F_0 &= \pm\sqrt{2m(E-V)} \equiv \pm p \\
G_1 &= - \frac{1}{4} [\log(E-V)]^\prime \ ,
\end{align*}
so that the wavefunction takes the form
\begin{align}
\label{wf_allow}
\Psi = \alpha_L \frac{1}{\sqrt{p}} \exp \left[ \frac{i}{\hbar} \int^{\lambda}_{\lambda_0}  p(l) dl \right] + \alpha_R \frac{1}{\sqrt{p}} \exp \left[ - \frac{i}{\hbar} \int^\lambda_{\lambda_0} p(l) dl \right]
\end{align}
where the coefficients $\alpha_L$ and $\alpha_R$ depend on the choice of the integration limit $\lambda_0$. In the classically forbidden region, the same reasoning leads to 
\begin{align}
\label{wf_forb}
\Psi = \alpha_+ \frac{1}{\sqrt{\tilde{p}}} \exp \left[ \frac{1}{\hbar} \int^{\lambda}_{\lambda_0}  \tilde{p}(l) dl \right]  + \alpha_- \frac{1}{\sqrt{\tilde{p}}} \exp \left[ - \frac{1}{\hbar} \int^\lambda_{\lambda_0} \tilde{p}(l) dl \right] \ ,
\end{align}
where $\tilde{p} = \sqrt{-2m (E-V)}$. Matching between both regime cannot be done immediately within the WKB approximation since it breaks down near the classical turning points where $V \rightarrow E$. One then again solves the Schr\"odinger equation but this time at the vicinity of the turning points $\lambda_\pm$. Around these points, the potential can be linearised and the Schr\"odinger equation reduces to an Airy equation whose solutions asymptotics in $-\infty$ and $+\infty$ are known. By matching these asymptotic form with the WKB solutions in both the allowed and forbidden regions, one obtains the so-called  connection formula. Crucially, these formula only modifies the real part of the wave function. They are thus critical in accounting for interference phenomena like resonant tunneling, but can be neglected while focusing on the tunneling exponent as we do in this work.

This formalism has been extended in~\cite{Bowcock:1991dr} to the multidimensional case. In this case, the Heisenberg equation using the WKB approximation becomes
\begin{align}
\label{eq:multidimSE}
& ( \nabla G)^2 - (\nabla F )^2  + \hbar \nabla^2 G= 2 m (E-U)  \\[1.1em]
& 2 \nabla F \cdot\nabla G = \hbar  \nabla^2 F \ .
\end{align}
where the last equation should be understood as the requirement that the divergence of the probability current $e^{2G/\hbar} \nabla F$ vanishes. We can see $\nabla S = \nabla F + i \nabla G$ as the ``momentum'' of the wave function.

In the semi-classical limit we are interested in, we want to neglect the $\hbar$ terms in the above equations. This implies the two conditions,
\begin{align}
(\nabla F )^2 &\gg \hbar \nabla^2 F  \\
(\nabla G )^2 &\gg \hbar \nabla^2 G \nn \  ,
\end{align}
which are broken either when $\nabla G ,\nabla F$ vanish (corresponding to the usual case $U-E = 0$) or when $\nabla^2 G, \nabla^2 F$ become very large. The latter occurs at caustic of the F-lines and G-lines, namely when initially neighbouring lines cross. For a slowly varying barrier compared to its steepness, this happens at the turning point of F-lines. We will be interested to the case of tunneling with an initial transverse momentum, so that $\nabla F$ will be non-vanishing parallel to the barrier. In that case, the matching with the quantum regime will occur along the caustic, or in our approximations, when the momentum is perpendicular to the barrier (for an almost step-function, this is simply at the barrier itself).
Note that one can also find an interpolation at the boundary between classical and quantum regime in terms of Airy functions, see~\cite{Bowcock:1991dr}.

We are left with solving~\eqref{eq:multidimSE} under the barrier. Bowcock and Gregory described a step by step procedure allowing to solve this system perturbatively assuming,
\begin{align}
\hbar \ll \frac{E}{U},\frac{(\nabla F)^2}{U} \ll 1 \ .
\end{align}
Indeed at zeroth order, Eq.~\eqref{eq:multidimSE} reduces to the standard form 
\begin{align}
\label{eq:SE_zeroth}
( \nabla G)^2  = 2 m U
\end{align}
which can be solved by using the momentum transfer equation derived from~\eqref{eq:SE_zeroth},
\begin{align}
\nabla G_{\nabla G}  = m \nabla U 
\end{align}
which amounts to search for the integral lines of the gradient of $G$. Once these lines have been found, we can use the second equation of~\eqref{eq:multidimSE} to prove that $F$ is constant along such lines, and subsequently find $\nabla F$ under the barrier. Replacing in the first equations of~\eqref{eq:multidimSE} leads to the full equation for $G$ at first order
\begin{align}
\label{eq:SE_first}
( \nabla G)^2  = 2 m U - \big(  2 m E - ( \nabla F)^2   \big) \ .
\end{align}
Solving step-by-step, one can obtain the tunneling rate.

\section{One-dimensional reduction of the FSE}
\label{app:1dRed}

If one assumes that the hypersurface $\scr{H}$ is one-dimensional, the problem can be solved completely in the WKB approximation. We review in this appendix this case, following~\cite{Bitar:1978vx,copeland_no_2008}, which can be used to obtain the vacuum-to-vacuum tunneling rate. Since the initial state is time-independent, we can start our analysis directly from the FSE for an  eigenstate of energy $E$,
\begin{align}
\label{eq:FSE2bis}
& \intv dx^3 \left[ \left(\fd{F}{\phi}\right)^2 - \left(\fd{G}{\phi}\right)^2 + \hbar \fdd{G}{\phi} \right] = 2 E - 2 U(\phi) \\
& \intv dx^3 \left[ \fd{F}{\phi}  \fd{G}{\phi} - \hbar \fdd{F}{\phi} \right] = 0 \ \nn,
\end{align}
where we have omitted the space dependence of the functional derivative for notational simplicity. The one-dimensional $\scr{H}$ corresponds to the field configurations $\phi_s(\lambda)$, such that 
\begin{align}
\label{eq:1DEoM}
\displaystyle \left.\frac{\delta }{\delta \phi_\perp }\Psi(\phi) \right|_\scr{H} =  0  \qquad \Rightarrow  \qquad \begin{cases}
\displaystyle \left.\frac{\delta F}{\delta \phi_\perp } \right|_\scr{H} =  0  \\
\displaystyle \left.\frac{\delta G}{\delta \phi_\perp } \right|_\scr{H} =  0 
\end{cases}\ .
\end{align}
We can write the functional derivative along the line $\scr{H}$ by decomposing it as:\footnote{This is easily understood by going back to the finite dimensional limit where the previous formula can be written as,
	\begin{align*}
	\vec{\nabla} = \frac{\vec{n}_{\parallel 1}}{| \vec{n}_{\parallel 1}|^2} \d_{\vec{n}_{\parallel 1}} + (\perp \text{ field configurations}).
	\end{align*}
}
\begin{align*}
\left. \fd{}{\phi}\right|_\scr{H}  &= \frac{\d_\lambda \phi_s}{ \mlam} \intv dx \d_\lambda \phi_s \left. \fd{}{\phi}\right|_\scr{H}  +  \displaystyle \left.\frac{\delta G}{\delta \phi_\perp } \right|_\scr{H} \\
&\equiv \frac{\d_\lambda \phi_s}{ \mlam} \d_\lambda  +  \displaystyle \left.\frac{\delta G}{\delta \phi_\perp } \right|_\scr{H} \ ,
\end{align*}
where we have introduced the field normalisation,
\begin{align}
\label{eq:mass1d}
\mlam &\equiv \intv dx^3 (\d_\lambda \phi_s)^2  \ .
\end{align}
Using this decomposition of the functional derivative on the hypersurface $\scr{H}$, we can then reduce the FSE Eq.~\eqref{eq:FSE2} to the system of ordinary differential equations in $\lambda$, 
\begin{align}
\label{eq:systFSE1d}
&(\d_\lambda F)^2 - (\d_\lambda G)^2  + \hbar~ \d_\lambda^2 G  = -2 \mlam \left(U(\phi)-E \right) \\
& \d_\lambda F \d_\lambda G - \hbar ~\d_\lambda^2 F  = 0 \ \nn.
\end{align}
Following the WKB approximation, we will be interested in the tunneling process itself in the semi-classical limit, meaning that we suppose
\begin{align*}
\hbar ~\d_\lambda^2 (F+iG) \ll  (\d_\lambda (F+iG))^2 \ .
\end{align*}

In the semi-classical approximation, the system~\eqref{eq:systFSE1d} has two regimes depending on the sign of $U(\lambda) - E$.
The case $U - E < 0$ corresponds to the classical region. In this regime, we find 
\begin{align}
\d_\lambda F &= \pm \frac{1}{\hbar}\sqrt{2 \mlam (E-U)} + \scr{O}(1) ~\equiv~ \pm \frac{i}{\hbar} p + \scr{O}(1)\\
\d_\lambda G &=  \frac{1}{4}  \d_\lambda  [\log(E-U)] + \scr{O}(\hbar) \nn \ ,
\end{align}
along with the ``energy conservation'' relation
\begin{align}
\mlam = 2 \left(E- U (\lambda) \right) \ .
\end{align}
The fact that the previous equation refers to energy conservation can be readily seen by using the definition of $\mlam$ in Eq.~\eqref{eq:mass1d} and recasting it as
\begin{align}
\label{eq:1dEcons_class}
\intv dx^3 \left( \frac{(\nabla \phi )^2}{2} + V(\phi) + \frac{1}{2} \left( \frac{\d \phi}{\d t} \right)^2 \right) = E \ .
\end{align}

In the quantum region, which corresponds to $U - E > 0$, we obtain:
\begin{align}
\label{eq:Gsol1d}
\d_\lambda G &= \pm \frac{1}{\hbar}\sqrt{2 \mlam (U-E)} + \scr{O}(1) ~\equiv~ \pm \frac{i}{\hbar} p + \scr{O}(1)\\
\d_\lambda F &= - \frac{1}{4} \d_\lambda  [\log(U-E)]+ \scr{O}(\hbar) \nn \ ,
\end{align}
and 
\begin{align}
\label{eq:1dEcons}
\mlam = 2 \left(U (\lambda) - E\right) \ .
\end{align}
The wave functional can then be written in the standard form (in the quantum region) as function of two constants $\alpha_+$ and $\alpha_-$,
\begin{align}
\Psi (\lambda) = \frac{1}{\sqrt{p}} \left[ \alpha_+ \exp \left( \frac{1}{\hbar}  \int_{\lambda_i}^\lambda dy p(y) \right)   + \alpha_-  \exp \left(- \frac{1}{\hbar}  \int_{\lambda_i}^\lambda dy p(y) \right) \right].
\end{align}

Contrary to the Quantum Mechanical case, we are however not done yet, since we still do not have any information on the form of the path $\phi_s$. Using the explicit form for the wave functional in the WKB approximation, we can recast the second equation of~\eqref{eq:FSE2} in a solvable form. Writing  $\delta \phi_\perp$ an infinitesimal field variation orthogonal to $\phi_s$
\begin{align*}
\displaystyle \left.\frac{\delta }{\delta \phi_\perp }\Psi(\phi) \right|_{\phi_s} &=  0  &\Longleftrightarrow&& \displaystyle \intv dx^3 \left(\delta \phi_\perp (x) \left.\frac{\delta }{\delta \phi} \Psi(\phi) \right|_{\phi_s} \right) &= 0 \\
&&\Longleftrightarrow&& \displaystyle \intv dx^3 \left(\delta \phi_\perp (x) \left.\frac{\delta }{\delta \phi} (F + i G) \right|_{\phi_s} \right) &= 0 \ .
\end{align*}
We can now use the explicit form of our action to express this last equation as a simple equation of motion for $\phi_s$. Let us first focus on the quantum regime. We have $F$ constant and $G = \int_{\lambda_i}^{\lambda_f} d \lambda \sqrt{ 2 \mlam  (U - E )} $. Replacing in the previous equation, we find 
\begin{align*}
\int dx^3 \delta \phi_\perp (x) \left( \int_{\lambda_i}^{\lambda_f} d \lambda \sqrt{ 2 \mlam  (U - E ) }  \right) &= 0 
\end{align*}
and finally
\begin{align}
&\int dx^3 \delta \phi_\perp (x) \int_{\lambda_i}^{\lambda_f} d \lambda \left( \sqrt{2 \frac{(U - E )}{\mlam}} \delta{ \mlam }+\sqrt{\frac{\mlam}{(U - E )}} \delta{U} \right) = 0 \nn \\ \label{eq:EoM1d}
& \Rightarrow \qquad  \frac{\d^2}{\d \lambda^2}\phi_s + \Delta \phi_s + \frac{\d  V}{\d \phi}   = 0  \ .
\end{align}
We therefore recover the standard equations of motion for the field in Euclidean time. Coupled to the solution for the wave functional~\eqref{eq:Gsol1d}, this procedure has been used in~\cite{Bitar:1978vx} to recover the tunneling rate from a false vacuum to a true vacuum by using the Coleman-De Luccia instanton as a solution of~\eqref{eq:EoM1d}. The classical case can be treated completely similarly, and one would recover the equation of motion for the field in real time. An important comment is that the ``time'' parameter $\lambda$ defined along the curve $\scr{H}$ and satisfying Eq.~\eqref{eq:1dEcons_class} is different from the time $t$ which gives the evolution of the wave functional. In the calculation, $\lambda$ simply appears as a convenient parametrisation of the classical path, for which the equation of motions take their standard form.

\section{Successive tunneling events}
\label{sec:Twotunneling}

An interesting consequence of the formalism introduced in Sec.~\ref{Sec:multiRed} is that it can be adapted to described the case of multi-tunneling events. Let us focus on the simplest case where the first tunneling event does not interact with the second one. It is obvious that the tunneling rate for such double-bubble emergence will simply be the product of the tunneling rate for each event. Nonetheless, the traditional instantonic formalism can not rigorously describe it since the initial state for the second tunneling event is time-dependent due to the first bubble growth. 

If the two events are spatially decorrelated (in the sense that the field from the outer bubble is constant on the wall of the inner one), we can write the effective potential as,
\begin{align}
U(t,\lambda) = U_{g}(t) + U_{\rm tun}(\lambda) \ .
\end{align}
where $U_{g}$ is the effective potential of a single classically growing bubble and $U_{\rm tun}(\lambda)$ is the effective potential of a single tunneling bubble. We can use a field configuration described by,
\begin{align}
\label{eq:2bubbbles}
\phi(\vec{x},t,\lambda) = \phi_{g1} (\vec{x},t) + \phi_{\rm tun} (\vec{x},\lambda) \ , 
\end{align}
where the first term $\phi_{g1}$ describe the growing bubble and $\phi_{\rm tun}$ the second tunneling event as a solution of the equation of motion Eq.~\eqref{eq:EoM_full}. The phase of the wave functional before tunneling is then simply given by $F=F_{g1}(t)$ where $F_{g1}(t)$ is the WKB solution for the growing bubble. Since the effective potential is separated, the FSE Eq.~\eqref{eq:FSE3} is trivially satisfied separately for its classical part
\begin{align}
- \d^i F \d_i F = 2 (U_{g1}-E) \ ,
\end{align}
and its quantum one
\begin{align}
&\d^i G \d_i G  = 2 U_{\rm tun} \ ,   
\end{align}
as long as $G$ is given by the standard one bubble tunneling expression, and $F=F_{g1}(t)$.

Overall the tunneling rate of the two events is then the sum of both rates as expected. 

\newpage

\bibliographystyle{utphys}
\bibliography{OscVacTheory}

\end{document}